%% file: main.tex
\documentclass[lettersize,journal]{IEEEtran}
\usepackage{amsmath,amsfonts}
\usepackage{array}
\usepackage[caption=false,font=small,labelfont=sf,textfont=sf]{subfig}
\usepackage[font=small,labelfont=bf]{caption}
\usepackage{textcomp}
\usepackage{stfloats}
\usepackage{float}
\usepackage{url}
\usepackage{verbatim}
\usepackage{graphicx}
\usepackage{cite}
\usepackage{balance}
\usepackage{algorithm}
\usepackage{algpseudocode}
\usepackage{enumitem}
\usepackage[english]{babel}
\usepackage{makecell}
\usepackage{hyperref}
\addto\extrasenglish{%
}

\graphicspath{{figs/}{figures/}{pictures/}{images/}{./}} 
\usepackage[svgnames,dvipsnames]{xcolor}
\definecolor{ride_hailing_technology}{RGB}{242, 142, 44}
\definecolor{middle_east_threats}{RGB}{89, 161, 79}
\definecolor{controversies_and_challenges}{RGB}{78, 121, 167}
\definecolor{criminal_justice}{RGB}{153, 86, 51}
\definecolor{political_protests}{RGB}{130, 48, 207}
\definecolor{concerns_and_controversies}{RGB}{131, 230, 77}
\definecolor{presidential_election_and_controversies}{RGB}{48, 202, 129}
\definecolor{presidential_eligibility}{RGB}{240, 98, 130}
\definecolor{ben_carson}{RGB}{187, 187, 187}
\definecolor{vis_web_data}{RGB}{240, 129, 55}

\begin{document}
\title{HINTs: Sensemaking on large collections of documents with \textbf{H}ypergraph visualization and \textbf{INT}elligent agents}
\author{
Sam Yu-Te Lee, Kwan-Liu Ma, \IEEEmembership{Fellow, IEEE}
\thanks{Sam Yu-Te Lee and Kwan-Liu Ma are with the Department of Computer
Science, University of California at Davis, One Shields Ave., Davis, California
95616. E-mail: {ytlee, klma}@ucdavis.edu}
}

\markboth{IEEE TRANSACTIONS ON VISUALIZATION AND COMPUTER GRAPHICS}
{Sam Yu-Te Lee, Kwan-Liu Ma, \MakeLowerCase{\textit(et al.)}: HINTs: Sensemaking on large collections of documents with \textbf{H}ypergraph visualization and \textbf{INT}elligent agents}%

\maketitle

\begin{abstract}
Sensemaking on a large collection of documents (corpus) is a challenging task often found in fields such as market research, legal studies, intelligence analysis, political science, computational linguistics, etc.
Previous works approach this problem either from a topic- or entity-based perspective, but they lack interpretability and trust due to poor model alignment.
In this paper, we present HINTs, a visual analytics approach that combines topic- and entity-based techniques seamlessly and integrates Large Language Models (LLMs) as both a general NLP task solver and an intelligent agent.
By leveraging the extraction capability of LLMs in the data preparation stage, we model the corpus as a hypergraph that matches the user's mental model when making sense of the corpus. The constructed hypergraph is hierarchically organized with an agglomerative clustering algorithm by combining semantic and connectivity similarity.
The system further integrates an LLM-based intelligent chatbot agent in the interface to facilitate sensemaking.
To demonstrate the generalizability and effectiveness of the HINTs system, we present two case studies on different domains and a comparative user study.
We report our insights on the behavior patterns and challenges when intelligent agents are used to facilitate sensemaking.
We find that while intelligent agents can address many challenges in sensemaking, the visual hints that visualizations provide are necessary to address the new problems brought by intelligent agents.
We discuss limitations and future work for combining interactive visualization and LLMs more profoundly to better support corpus analysis.
\end{abstract}

\begin{IEEEkeywords}
Text visualization, sensemaking, hypergraph, hierarchical clusters, corpus analysis, large language models
\end{IEEEkeywords}

\input{01_Introduction}
\input{02_Related_Works.tex}

\input{03_Design_Rationale.tex}

\input{04_Methodology.tex}

\input{05_Visualization.tex}

\input{06_System_Design.tex}

\input{07_Usage_Scenario.tex}

\input{08_User_Study.tex}

\input{09_Limitations_and_Future_Work.tex}

\input{10_Conclusion.tex}

\subsection{Acknowledgments}
\noindent This research is supported in part by the National Science Foundation under Grant No. IIS-2427770.

\bibliographystyle{abbrv-doi-hyperref}

\bibliography{template}

\vspace*{-7cm}
\begin{IEEEbiography}[{\includegraphics
[width=1in,height=1.25in,clip,
keepaspectratio]{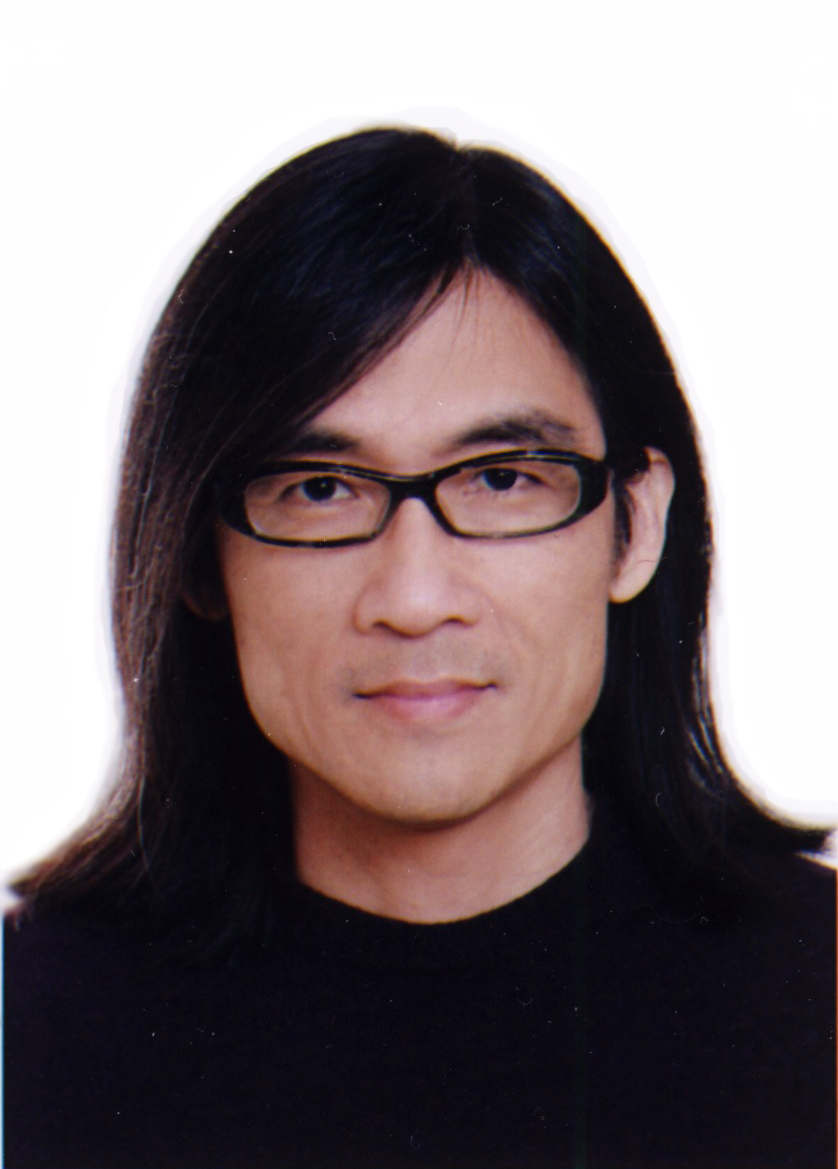}}]
{Kwan-Liu Ma} is a distinguished professor of computer science at the University of California,
Davis, where he leads VIDI Research Group. Professor Ma received his PhD degree in 
computer science from the University of Utah in 1993. His research interests include visualization, computer graphics, human computer interaction, and machine learning. For his significant 
research accomplishments, Professor Ma has received many recognitions, among others 
the NSF PECASE award in 2000, IEEE Fellow 2012, the IEEE VGTC Visualization Technical
Achievement Award in 2013. the IEEE Visualization Academy in 2019, and ACM Fellow in 2024.
He has served as papers co-chair for SciVis, InfoVis, EuroVis, PacificVis, and Graph Drawing,
and on the editorial board of IEEE TVCG (2007-2011) and IEEE CG\&A (2007-2019). 
Professor Ma presently serves on the editorial boards of the Journal of Visual Informatics, 
and the Journal of Computational Visual Media, and ACM Transactions on Interactive Intelligent Systems. Contact him via email: ma@cs.ucdavis.edu.
and research and other interests.
\end{IEEEbiography}

\vspace*{-6cm}
\begin{IEEEbiography}[{\includegraphics
[width=1in,height=1.25in,clip,
keepaspectratio]{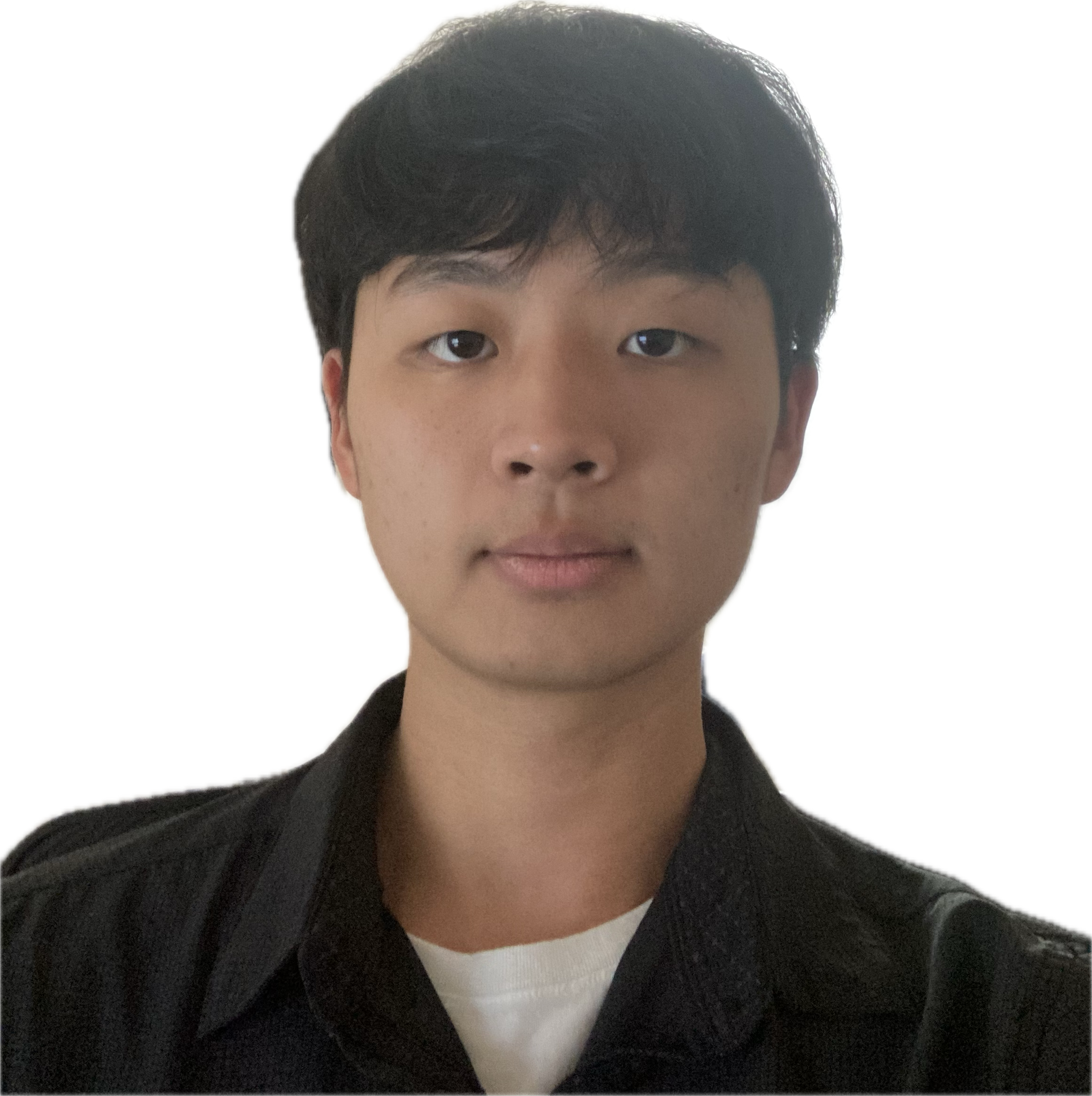}}]{Sam Yu-Te Lee}
received a BS degree in computer science from Nanjing
University in 2020. He is currently a 
PhD student in computer
science at the University of California, Davis. His research interests include text data visualization and analytics.
\end{IEEEbiography}
\clearpage
\appendix
\input{Prompts}

\end{document}

%% file: 01_Introduction.tex
\section{Introduction}
\noindent \IEEEPARstart{T}{ext} data is ubiquitous.
From news articles and social media posts to scientific publications, the tremendous amount of text data that is produced poses not only opportunities but also a great challenge to anyone who needs to analyze them.
Visual Analytics (VA) mitigates this challenge by combining mathematical models and visualizations to automate the sensemaking process and reduce the cognitive load.
\textit{Model Alignment}~\cite{chuang2012interpretation}, the alignment of analysis tasks, visual encodings, and model choices,
greatly affects users' interpretation and trust in visual analytic systems.
In modern text analysis, the \textit{models} that affect visual analytics system design is often replaced with the Natural Language Processing (NLP) tasks that Deep NLP models (e.g., BERT~\cite{devlin2018bert}) can perform, such as named entity recognition, topic modeling, or sentiment analysis. 
However, these NLP tasks often align poorly with users' analysis tasks.
For example, topic models are commonly used to model the topical structure of text documents.
Most topic models characterize \textit{topic} as a probabilistic distribution spanning a given vocabulary~\cite{vayansky2020review}.
This transformation from a \textit{topic}, an apprehensible concept, to a \textit{probabilistic distribution}, a mathematical concept that models can operate on, hinders proper model alignment.
The misalignment between NLP tasks and user tasks limits the usage of visual analytics systems for users who are not familiar with the underlying models.

Recent advances in large language models (LLMs) present a promising solution to this problem.
LLMs have proven their performance as intelligent agents in open-context question-answering (prompting) due to their strong capability to understand user intent~\cite{brown2020language}.
Visualization researchers have adopted them to assist data transformation~\cite{wang2023dataformulator} or directly generate visualization~\cite{maddigan2023chat2vis}, in which LLMs are used as ``Agents'' that delegate some tasks.
However, applying LLMs in the data-preparation stage of the visualization pipeline is under-explored.
Topics~\cite{atzberger2023evaluatetopicmodel}, sentiments~\cite{beasley2021through}, concepts and entities~\cite{park2018conceptvector,cao2010facetatlas} are common analysis targets in text analysis that require a dedicated data preparation stage. Researchers have been using NLP models to extract them, but by using LLMs as a general-purpose NLP task solver, visualization creators can now extract more than just topics, sentiments, or entities.
With prompting, visualization creators can extract precisely what they want to visualize from the documents. It also enables them to provide domain-specific contexts for the extraction, easily achieving human-level extraction results automatically. 
In the previous example, instead of relying on abstruse and unfathomable probabilistic models, visualization creators can process the text data and summarize the topics of the documents using a prompt.
A visualization creator can ask in a prompt:\@ \textit{``What are the topics of these articles?''}, and the model would give a human-like response, such as \textit{``The articles are about Covid-19''}.
If there is a pre-defined list of topics in mind, it can be added directly in the prompt and the model can assign topics accordingly.

Such advance in data processing capabilities opens many doors for visualization creators. 
In this work, we present HINTs (\textbf{H}ypergraph visualization and \textbf{INT}elligent agents), a VA system that combines LLM's ability both as ``Agents'' and as general NLP task solvers to help users make sense of a large corpus.
The HINTs system models a corpus as a hypergraph, where the nodes are documents and salient keywords.
Once the hypergraph is constructed, it is then hierarchically clustered and visualized with enhanced space-filling curve layouts~\cite{muelder2008sfc}.
We showcase how LLMs are used in the data preparation stage to take into account the user analysis tasks and visual encodings.
The system supports interactive exploration and reorganization of the documents, after which users can use an intelligent chatbot agent to conduct further analysis of the documents.
To the best of our knowledge, no published visual analytics system has adopted LLMs for Model Alignment.
Using the system, we demonstrate two different ways that integrate LLMs into a VA system and advocate for their incorporation into other parts of the visualization pipeline.

The contributions of our work are as follows:
\vspace*{-0.1cm}
\begin{itemize}
  \setlength\itemsep{0em}
    \item We introduce an LLM-based data preparation pipeline that is capable of extracting topics and salient keywords from a given corpus for better model alignment.
    \item We extend space-filling curve layouts to visualize hierarchical clusters in large hypergraphs interactively. 
    \item We develop a novel VA system, HINTs, that facilitates users to make sense of a corpus. From the user study of HINTs, we report our observations on user behavior patterns and insights on the benefit of visualization.
\end{itemize}

%% file: 02_Related_Works.tex
\vspace*{-0.5cm}
\section{Related Works}
\noindent Our work builds upon existing visual analytic systems for document intelligence and advances in the application of LLMs. We first give a breakdown of the data preparation approaches in existing visual analytic works, then switch focus to the visual designs in those works. Finally, we introduce existing advances in LLMs in different application scenarios, with a focus on using them as a general NLP task solver. 
\vspace*{-0.5cm}
\subsection{Data Preparation for Large Collections of Text}
\subsubsection{Topic-based approaches}
Topic-based approaches employ variations of topic models to organize the documents.
Each topic is often presented as a ``bag of words'', which can be in the form of a sequence of words~\cite{yang2017vistopic, cho2016vairoma, dou2013hierarchicaltopics,han2022hisva, alexander2014serendip, lee2012ivisclustering} or word clouds~\cite{oelke2014topiccoin, cho2016vairoma}.
The modeling result is used as an overview of the corpus for subsequent analysis tasks. 
Despite their popularity, using topic modeling results as an overview, as well as its ``bag-of-word'' representation, is reported by Lee et al.~\cite{lee2017humantopicmodel} to be problematic, especially for non-expert users.
Chuang et al.~\cite{chuang2012interpretation} concluded that these problems arise from a misalignment between the analysis task, visual encoding, and model.
Sensemaking is challenging without a basic understanding of the model because the ``bag-of-words'' representation diverges from the user's mental model of a topic.
This misalignment limits the usage of topic models for non-expert users and makes the system prone to producing false insights. 

\subsubsection{Entity-based approaches}
A line of work that makes successful model alignments is the entity-based approach.
`Entities' usually include named entities (people, organizations, locations), or meaningful concepts known to an existing knowledge base.
The earliest of such approaches is Jigsaw~\cite{Stasko2007jigasw}, where entities are linked if they appear in the same document.
FacetAtlas~\cite{cao2010facetatlas} generalizes the idea of entity to `facets' which can be entities or any keywords or user's interest.
ConceptVector~\cite{park2018conceptvector} uses `concept' to represent a similar idea. 
Generally, entity-based approaches exhibit better model alignments than topic-based approaches~\cite{chuang2012interpretation}, 
but the polysemy of natural language makes them prone to produce false positives~\cite{park2018conceptvector}.
In our work, we use \textit{keywords} to represent entities or concepts that appear in the documents.
Previous works use computational metrics such as TF-IDF or \textit{saliency}~\cite{chuang2012termite} to extract keywords.
However, computational metrics often cannot encode high-level semantics very well, causing them to sometimes unalign with human judgments.
Instead, we rely on LLMs' ability to understand natural language to identify significant keywords from unstructured texts.
We ensure that the keywords are salient by exploiting LLM's ability to understand semantic contexts.

\subsubsection{Embedding-based approaches}
Finally, an important line of work organizes documents by directly modeling their semantic similarity~\cite{steinbach2000doccluster}.
Documents are first projected into a high-dimensional vector space where similarity can be measured, and then a dimensionality reduction technique (e.g., t-SNE) is used to project the documents back into a two-dimensional space for visualization.
Earlier works construct a sparse vector using term-frequency based scores such as \textit{TF-IDF} or BM25~\cite{choo2013utopian,sherkat2018interactive}.
More recently, the success of pre-trained language models like BERT~\cite{devlin2018bert} popularizes the idea of embedding documents in a dense vector space~\cite{narechania2022vitality,tu2023sdrquerier,qiu2022docflow}.
The embedding can then be used for document retrieval~\cite{karpukhin-etal-2020-dense, izacard2022unsupervised} or visualization.
Embedding-based approaches also exhibit proper model alignment, as the semantic distance in vector space fits the user's mental model in the analysis task (i.e., finding similar documents). 
However, the result often lacks explainability and prevents users from trusting the result.
Recently, Raval et al.\cite{raval2023explainandtrust} proposed the use of LLMs to provide explainability to embeddings-mappings visualizations.
We adopt a similar approach in our system to provide interpretability to the clustering result.

\vspace*{-0.35cm}
\subsection{Visual Interfaces for Corpus Analysis}
In this section, we review visual interfaces designed for the exploration or analysis of a corpus. Even though there are some overlaps with the previous section, here we focus on their visual representations.

Since the most prominent technique for organizing a corpus is topic modeling, many of the visual representations are centered around the topic modeling results, most of them involving topic co-occurrences or hierarchies.
Serendip~\cite{alexander2014serendip} uses a reorderable matrix of documents and topics for the exploration of topic occurrences and their importance.
HierarchicalTopics~\cite{dou2013hierarchicaltopics} uses a tree visualization where each node is a topic to visualize the topic hierarchies.
Due to the topic model's ``bag-of-word'' representation, many previous works use word clouds in their visualization.
SolarMap and FacetAtals~\cite{cao2011solarmap, cao2010facetatlas} use a radial word cloud for word appearances in topics.
VisTopic~\cite{yang2017vistopic} combines a sunburst diagram and a modified tag cloud to visualize the topic hierarchy and word occurrences at the same time.
An exception is PhraseMap~\cite{tu2022phrasemap}, where a map metaphor is used to visualize the extracted technical phrases. Similar to our approach, they use a Gosper curve to create the layout. Different from topic- and entity-based approaches, embedding-based approaches typically employ dimensionality reduction techniques and use a scatter plot as their primary visualization~\cite{park2018conceptvector, caillou2020cartolabe, narechania2021vitality, choo2013utopian}. 

One key distinction of the visualization between topic-based and embedding-based approaches is the unit of abstractions. 
In topic-based approaches, the unit of abstractions is ``bag-of-words'' (topics) and their hierarchies. 
In embedding-based approaches, the unit of abstraction is the documents. 
We note that for a better model alignment in the context of corpus sensemaking and analysis, the unit of abstraction must be the documents since the unit of analysis is the document.
Further discussion on this matter is presented in~\autoref{sec: design_rationale}.

\vspace*{-0.5cm}
\subsection{LLMs as General NLP Task Solvers}
LLMs have been proven to excel at many NLP tasks, such as sentiment analysis or summarization, and even more complicated tasks that in the past could only be done by humans. 
Gilardi et al.~\cite{gilardi2023chatgpt} reported that ChatGPT can generate annotations that outperform crowd-workers on commercial platforms by about 25 percent accuracy on average, and cost over thirty times less financially.
Beyond simple annotations, Wu et al.~\cite{wu2023llms} shows that LLMs can replace multi-stage crowdsourcing pipelines with careful prompt designs.
The NLP tasks closest to our work involve Information Extraction (IE), a line of NLP tasks that aims to identify structured information of interest from unstructured text.
Some of its subtasks include Named Entity Recognition (NER), Relation Extraction (RE) and Event Extraction (EE)~\cite{nasar2021named, xiang2019surveyee}.
Li et al.~\cite{li2023evaluateChatgpt} comprehensively evaluated the capabilities of ChatGPT for common IE tasks.
They found that the model excels under the Open-IE setting, where the model directly extracts information from documents but performs poorly under the Standard-IE setting, where the model chooses a correct label from candidate labels.
Similarly, Zhang et al.~\cite{zhang2023extractive} reported that ChatGPT performs poorly on extractive summarization, where sentences in the text are treated as ``candidates'' and the model is instructed to select sentences to form summarization.
A common reason for the poor performance in these tasks is that they are essentially supervised learning tasks, tasks that the model is not trained to perform.
On the other hand, some works have shown the limitations of LLMs such as ``hallucination''~\cite{bang2023multitask} and ``faithfulness''~\cite{zhou2023contextfaithful} issue reported by existing works.
To make the best use of ChatGPT for NLP tasks, we are informed by these early experiments and carefully design our extraction tasks accordingly.

%% file: 03_Design_Rationale.tex
\begin{figure*}[htbp]
 \centering 
 \includegraphics[width=0.95\textwidth]{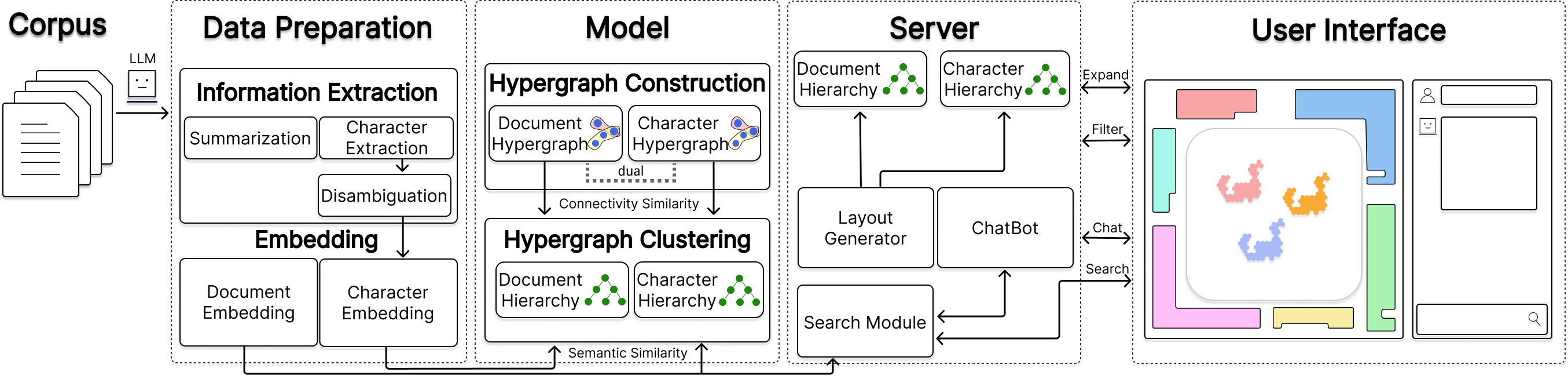}
 \caption{Data processing pipeline of HINTs. 
 Starting from a corpus of unstructured texts, each document goes through the data preparation stage to extract the salient keywords.
 Then the documents and keywords are both embedded into a vector space.
 The model stage constructs a document hypergraph and a keyword hypergraph, which are then clustered separately by combining connectivity similarity and semantic similarity.
 The clustered hypergraphs are hosted on the server and visualized in the user interface.
 Users can expand, filter, or search the hypergraphs to explore the corpus and select documents to be analyzed with a chatbot.
  }\label{fig:pipeline}
\vspace*{-0.5cm}
\end{figure*}

\vspace*{-0.5cm}
\section{Design Rationale}\label{sec: design_rationale}
\noindent The HINTs system is designed for experts and analysts to make sense of a corpus.
Our design rationale is based on the design guidelines proposed by Chuang et al.~\cite{chuang2012interpretation}.
We reuse their definitions of \textit{Model Alignment}, \textit{Progressive Disclosure}, and \textit{Unit of Analysis} when describing our design rationale.
We first identify common analysis tasks from previous works.
Then, we derive our design considerations (DC) from the analysis tasks.
We take the DCs into account during the data preparation and visualization design process in~\autoref{sec: methodology}.
\vspace*{-0.4cm}
\subsection{Analysis Tasks}\label{sec: analysis_tasks}
We derive the analysis tasks from topic- and entity-based approaches.
Topic-based approaches aim to support document understanding by visualizing the topic structure of the documents.
Entity-based approaches support investigative analysis by visualizing entity occurrences and relations.
We aim to support both tasks simultaneously as they are fundamental to subsequent tasks and intertwined in a real-world scenario.

We first generalize entities to \textit{Keywords}, which are the core entities or concepts that appear in the documents.
For example, in a news article, the keywords can be named entities in the title or protagonists of the news event.
In a research article, the keywords can be the proposed models or employed techniques.
This definition of keywords is generalizable enough for us to take user intent into account during the extraction.
Building around topics and keywords, our system designs visualizations and interactions to support the sensemaking of a corpus. 
Specifically, we support the following four analysis tasks:

\textit{T1: Understanding topic structures.} 
Topic-based approaches have shown that the topic structures of a corpus are essential to the sensemaking of the corpus. Each topic, consisting of a subset of documents, represents a significant semantic facet occurring in the corpus. By breaking down a corpus into topics, users break down the cognitively demanding sensemaking process into feasible subtasks in which they incrementally understand a single topic.

\textit{T2: Understanding keyword occurrences.}
Keywords also play an essential role in the sensemaking process, as shown by the entity-based approaches, so we treat them with equal importance to the topics. A subtle difference between keywords and topics is that keywords are carried by the documents, while topics carry the documents. Thus from the user's mental model, the keyword's occurrences are of interest, and the understanding of such occurrences contributes to the understanding of a document, a topic, and the whole corpus. 

\textit{T3: Navigating to documents of interest.}
A corpus, by its definition, contains too many documents for users to quickly find their target of interest. Through the understanding of topic structures and keyword occurrences, the target of interest is found and a subsequent step is to navigate to the target, either to verify or deepen existing understanding or resolve confusion. 
Although navigation does not directly contribute to sensemaking, it is an essential step to making progress.

\textit{T4: Understanding the content of a document and the meaning of a keyword.}
The most fine-grained level of sensemaking is to understand the content of a document and the meaning of a keyword. This can occur in cases where users need to get a clearer understanding of a topic or meet an unknown keyword that hinders a clear understanding. 
Previous works often ignore such a task or assume that the users already have enough knowledge to understand a topic or a keyword, which is questionable. We include this task to complete the sensemaking process and fill the gaps in existing research.

\vspace*{-0.2cm}
\subsection{Design Considerations}
To support the aforementioned analysis tasks, we derive the following design considerations (DCs):
\begin{itemize}
  \setlength\itemsep{0em}
  \item \textit{DC1: Overview of topic structures and keyword occurrences (\textit{T1, T2}).}
  Given a corpus, the topic structures and keyword occurrences can be complex and cover a wide range of documents and keywords. 
  The overview seeks to cover all the documents and keywords while hiding the details.
  This sets the ground for the user to discover their targets of interest.
  \item \textit{DC2: Progressive Disclosure (\textit{T1, T2, T3)}.}
  To facilitate incremental sensemaking, it is important to support users to drill down from a high-level overview to intermediate abstractions. 
  This includes disclosure of a specific topic's sub-structure, the containing documents, and the occurrences of keywords.
  It not only supports a better understanding of the topics and keywords but also directs users to find the target of interest. 
  \item \textit{DC3: Model Alignment (\textit{T1, T2}).}
  Having good interpretability for the presented topics and keywords is essential to support sensemaking, which is enabled by carefully considering Model Alignment. 
  Our choice of models should align well with the user's mental model when conducting the analysis tasks for better interpretability.
  This entails that our model should directly operate on the units of analysis (i.e., documents and keywords)
  \item \textit{DC4: Detailed analysis of the target of interest (\textit{T4}).}
  The investigation of topic structures and keyword occurrences often leads to a target of interest, which is a subset of documents.
  After such investigation, previous works usually only provide the user with a list of these documents.
  This is perhaps due to the lack of a unified way to analyze the target of interest under different contexts.
  The advance of LLMs presents a promising solution to this problem by transforming any analysis task into a question-answering task, enabling us to consider \textit{T4} in our system design. 
\end{itemize}

%% file: 04_Methodology.tex
\section{Methodology}\label{sec: methodology}
\noindent As shown in~\autoref{fig:pipeline}, starting from a corpus of unstructured texts,
we first use LLMs to extract the main keywords from each document.
The keywords are disambiguated and linked to a knowledge base if available.
We create and store document embeddings and keyword embeddings for further usage.

All LLM-based preprocessing used OpenAI's ``gpt-3.5-turbo-16k-0613'' model.
We provide all our prompts in supplemental material.
In the next stage, we construct a document hypergraph and a keyword hypergraph, which are then clustered separately by combining connectivity similarity and semantic similarity.
The clustering result is hosted on the server and visualized in an interactive user interface.
During the development of HINTs, we tested the pipeline on two datasets from different domains: a news article dataset (AllTheNews)~\cite{allthenews} and a visualization publication dataset (VisPub)~\cite{vispub}. 
In all the prompt-based extraction modules, we used two different versions of prompts to take the user's analysis tasks into account and guarantee optimal extraction results for the users.
Below, we describe each component in detail.
\vspace*{-0.4cm}
\subsection{Data Preparation}\label{sec: preprocessing}
The data preparation stage consists of five steps: summarization, keyword extraction, keyword disambiguation, document and keyword embedding, and topic label generation.
We use LLMs to perform all five steps as they outperform traditional approaches in accuracy and flexibility.
Additionally, we discuss how we design these tasks while taking the users' analysis tasks (\autoref{sec: analysis_tasks}) into consideration.
\subsubsection{\textit{Summarization}}\label{sec: summarization}
During sensemaking, users need a quick understanding of the content of the selected documents (\textit{T1}).  
Summarization helps condense the content into a concise form, making it easier for users to make sense of.
Zhang et al\cite{zhang2023extractive} reported that human users found the summaries generated by ChatGPT to be more interpretable and trustworthy than traditional approaches, making LLMs a better choice for our purpose.
We employed an instruction-based zero-shot prompt, where the model is instructed to act as a text summarizer.
The summarized documents are also used in subsequent data preparation pipelines for other prompts to better process and extract the most important information. 

\subsubsection{\textit{Keyword Extraction}}\label{sec: keyword_extraction}
We define keywords as \textit{phrases that are significantly discussed in the documents}.
We highlight the importance of ``significantly discussed'' here, as they serve as a unit of analysis in the user's tasks (\textit{T2}). 
The keywords are extracted with prompts, building upon the summaries generated in the previous step.
We use slightly adjusted versions of the description of a ``keyword'' in the prompts according to the dataset domain, making them even more tightly connected with the user's analysis tasks. 
For news articles, keywords are named entities or protagonists in the news event.
For research papers, keywords are models, techniques, or algorithms proposed or used by the paper.

In addition, we break down the process of keyword extraction into two steps to avoid the problem of extrinsic hallucination as discussed by Bang et al.~\cite{bang2023multitask}.
Below, we take keyword extraction in a news article dataset as an example. 
First, a sentence describing the most important event is generated from the summary. 
The event should be the primary focus of the news article, such as a geopolitical event like the G20 summit.
A few-shot prompt was adopted to give the LLM examples of what kind of events would be described as the main event. 
Next, all the main keywords involved in that event are extracted with another prompt, injecting the main event extracted in the previous step: 
\textit{``A major event is reported by a news article: \{event\_description\}, what are the main keywords that are majorly involved in the event?''.}
This provides the model with more context and yields better accuracy and lower chances of generating a hallucinated output.

\subsubsection{\textit{Keyword Disambiguation}}\label{sec: keyword_disambiguation}
Once the main keywords are extracted, a disambiguation step is done to create connections between documents (\textit{T2}).
In early experiments, we found that LLMs do not excel at disambiguating keywords.
We thus prioritize using an entity linking model~\cite{shen2014entity} to disambiguate keywords.
Existing models adopt a supervised learning approach which requires training data, so they are usually limited to a specific domain.
For datasets where an existing entity-linking model~\cite{ayoola2022refined} is available, we use it to disambiguate keywords.
Otherwise, we employ an embedding-based matching algorithm.
The algorithm is straightforward:
For every extracted keyword, a first prompt generates an explanation of the keyword in a few sentences. Then the explanations are embedded to create a pair-wise similarity matrix.
The pairs of keywords with a similarity score above a threshold are considered to be the same keyword.
Then a second prompt generates a unified title for the matched keywords to be displayed in the user interface.

\subsubsection{\textit{Document and Keyword Embedding}}\label{sec: embeddings} 
Embeddings are dense vectors that can be used to measure similarities between objects, which are often created on words, sentences, or documents in text analysis.
We adopted a similar approach here for both documents and keywords.
For documents, we embed the summaries generated in~\autoref{sec: summarization}.
For keywords that can be linked to an external knowledge base,
we embed the description of the keyword in the knowledge base.
Otherwise, we embed the explanation of the keyword generated with a simple prompt: \textit{``What is \{keyword\}?''}.
While previous works embed the keywords directly, we found that embedding the explanations produces more semantically precise results.
We used OpenAI's ``text-embedding-ada-002'' model for all embeddings,
allowing us to measure the semantic similarity between documents, keywords, and user queries in the same vector space.
The document and keyword embeddings are used subsequently to generate semantic clusters and visualized to support \textit{T1} and \textit{T2} as described in~\autoref{sec: clustering}.

\subsubsection{\textit{Topic Label Generation}}\label{sec: tag_assignment}
Later in the preprocessing pipeline, an agglomerative clustering algorithm is used to cluster the documents and keywords (\autoref{sec: clustering}). 
The algorithm outputs a hierarchy, where leaf nodes are documents or keywords and internal nodes are clusters.
However, the clusters are not readily interpretable for the users, hindering the support of \textit{T1}.
To address this, we assign human-readable labels to each cluster with prompts.
Our approach is similar to a recent work by Raval et al.~\cite{raval2023explainandtrust}, where LLMs are used to generate explanations for user-selected points.
However, our task is more complicated because the prompt needs to (1) consider the hierarchical structure, and (2) be able to process large clusters containing thousands of documents without breaking the token limit.

Our topic label generation follows a bottom-up approach:
(1) At the bottom level (one level above the leaf nodes), assign topic labels to the clusters using the document summaries.
The bottom-level clusters usually contain only a few documents, so the token limit is unlikely to be exceeded.
We insert the summaries generated in~\autoref{sec: summarization} into the prompt,
then instruct the LLM to generate a label composed of a single noun phrase for the given documents.
(2) At any level above, the cluster size increases quickly and the token limit is likely to be exceeded. We thus assign labels to the clusters using the labels of their children and randomly sampled document summaries.
This guarantees that the token limit will not be exceeded.
In early experiments. we found that the children's labels alone were not enough to generate a meaningful label for the parent cluster. 
To improve, we insert the document summaries randomly sampled from the cluster.
We enforce that each child has at least one article being sampled, and distribute the remaining token space proportionally to the size of each child.

\begin{figure}[t]
 \centering 
 \includegraphics[width=\columnwidth]{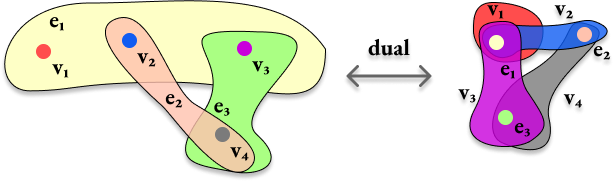}
 \caption{Illustration of the dual of a hypergraph. 
 Left: a hypergraph with 4 nodes $v_1, v_2, v_3, v_4$ and 3 hyperedges $e_1=(v_1, v_2, v_3), e_2=(v_2, v_3), e_3=(v_3, v_4)$. 
 Right: the dual of the hypergraph.
 Nodes in the dual hypergraph are now $e_1, e_2, e_3$ and hyperedges are $v_1=(e_1), v_2=(e_1, e_2), v_3=(e_1, e_3), v_4=(e_2, e_3)$.
 }
\vspace*{-0.5cm}
\label{fig: duality}
\end{figure}
\vspace*{-0.4cm}
\subsection{Models}

\subsubsection{\textit{Hypergraph Construction}}
A hypergraph is a generalization of a graph in which an edge can connect any number of nodes~\cite{fischer2021hypergraphsurvey}.
A hyperedge thus represents a multi-way relationship between nodes.
In our work, we model two types of hypergraphs: \textit{document hypergraph} and \textit{keyword hypergraph}.
In a document hypergraph, nodes are documents and hyperedges are keywords, and symmetrically in a keyword hypergraph, nodes are keywords and hyperedges are documents.
By modeling the corpus as hypergraphs, the system supports users to conduct both topic- and entity-based analysis simultaneously under a unified framework (\textit{DC1}).
Below, we explain the relation between these two types of hypergraphs and the construction process.

A hyperedge can be used to represent two types of multi-way relationships:
In a document hypergraph, a hyperedge (connecting document nodes) can be constructed between documents that mention the same keyword. 
In this case, the hyperedge represents the co-mention of a keyword.
In a keyword hypergraph, a hyperedge (connecting keyword nodes) can be constructed between keywords if they are mentioned together in the same document.
In this case, the hyperedge represents a co-occurrence relationship between keywords.
In other words, the two hypergraphs are built from the same corpus, using the same relationship information, but from different perspectives.
This allows us to utilize the \textit{dual} of a hypergraph to simplify the construction process.
The dual of a hypergraph is simply another hypergraph, where the nodes and hyperedges are interchanged, as shown in~\autoref{fig: duality}.
Using the duality feature, we first model the documents as nodes and keywords as hyperedges to construct the document hypergraph $H_D$.
The keyword hypergraph $H_C$ can then be easily constructed by taking the dual of $H_D$.
This construction process also allows us to map user interactions on documents and keywords to operations on a single hypergraph (\textit{DC1}).
Even better, the duality feature between documents and keywords matches the user's mental model (\textit{DC3}), making the visualization and interaction intuitive.

\subsubsection{\textit{Hierarchical Clustering}}\label{sec: clustering}
Once the two hypergraphs are constructed, we separately hierarchically cluster them using an agglomerative clustering algorithm.
Clusters in the document hypergraph represent topics that are discussed in the dataset.
Clusters in the keyword hypergraph represent keywords (entities or concepts) that are similar to each other.
For better interpretability of the clustering result, we further assign \textit{labels} to each cluster, as explained in~\autoref{sec: tag_assignment}.
Below, we describe our motivation for using an agglomerative clustering algorithm and its process.

Common clustering algorithms for graphs consider only graph connectivity.
However, for the best interpretability of the clustering result, the node embeddings must be also used in the clustering process to model semantic similarity.
We thus focus on discussing attributed node clustering algorithms.
Although there are existing approaches that can cluster attributed nodes on graphs such as EVA~\cite{citraro2020eva} and iLouvain~\cite{combe2015louvain}, they are not designed for hypergraphs.
Although Kamiński et al.~\cite{kaminski2021hgraphcommunity} have shown that it is feasible to generalize the modularity metric for graphs to hypergraphs (thus generalizing modularity-based clustering algorithms), we found in early experiments that it is not scalable and hard to incorporate node attributes.
Finally, we decided to follow the approach proposed by Kumar et al. ~\cite{kumar2020new}; i.e., transform the hypergraph into a graph and apply graph clustering algorithms. 
We decide to apply an agglomerative clustering algorithm~\cite{steinbach2000doccluster} to support progressive disclosure of the cluster structures (\textit{DC2}).

In agglomerative clustering, the key is to define node similarity and cluster similarity.
We can easily incorporate node attributes into the clustering process by defining the similarity between nodes and clusters as a weighted sum of attribute similarity $S_s$ and connectivity similarity $S_c$.
Since we're dealing with texts, we refer to the attribute similarity between nodes as semantic similarity. 
The semantic similarity $S_s(i, j)$ is the cosine similarity of the embeddings (dense vectors) of the two nodes, denoted as $vec_i$ and $vec_j$.
The connectivity similarity $S_c$ is the weighted Topological Overlap (wTO)~\cite{gysi2018wto},
which is a weighted generalization of the Overlap Coefficient~\cite{vijaymeena2016survey}, as shown in~\autoref{eq: connectivity_semantic_similarity}.
\begin{subequations}\label{eq: connectivity_semantic_similarity}
\begin{align}
    \mathbf{S_s}(i, j) = \frac{vec_i \cdot vec_j}{||vec_i|| \cdot ||vec_j||} \\
    \mathbf{S_c}(i, j) = \frac{\sum_{u=1}^N{w_{i,u}w_{u_j}} + w_{i,j}}{\min(k_i, k_j) + 1 - |w_{i,j}|}
\end{align}
\end{subequations}
where $k_i = \sum_{j=1}^N |w_{i,j}|$ is the total weight of the edges connected to node $i$.
Finally, a weighting factor $\alpha$ is used to balance the two similarities, as shown in~\autoref{eq: similarity}.
\begin{equation}\label{eq: similarity}
    \mathbf{S} = \alpha \mathbf{S_s} + (1-\alpha) \mathbf{S_c}
\end{equation}
For the similarity between clusters, we used centroid similarity, i.e., the similarity between two clusters is the similarity between the centroids of the two clusters.
The algorithm takes a hypergraph $H=(V, E)$ and the embeddings of each node as input, and outputs a sequence of partitions $P=P_1, P_2, \dots P_k$.
Each partition corresponds to a level in the hierarchy.
Pseudo-codes of the algorithm can be found in supplemental materials.

%% file: 05_Visualization.tex
\begin{figure*}[t]
  \centering
  \includegraphics[width=\textwidth, keepaspectratio]{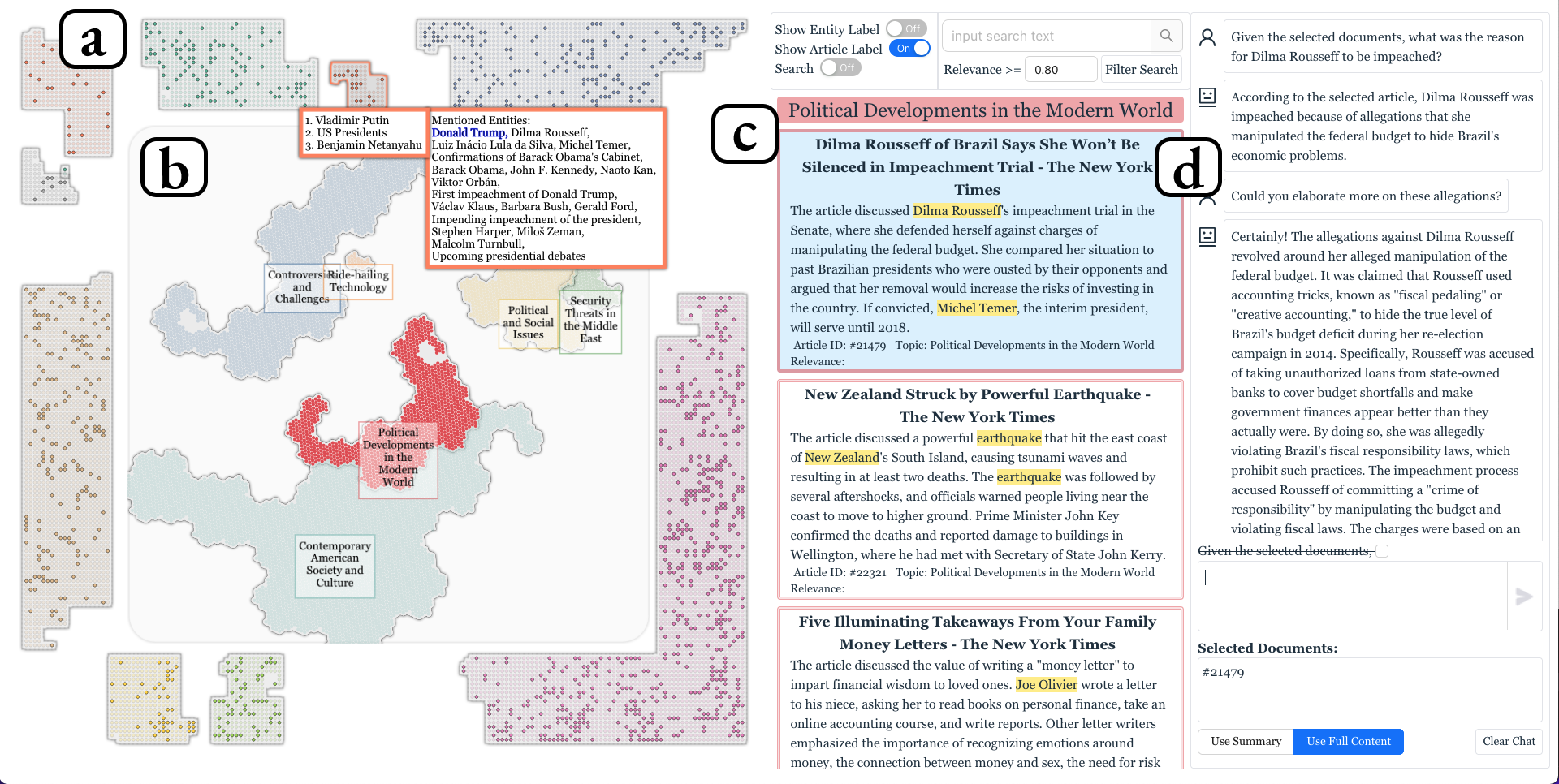}
  \caption{The HINTs system. 
  (a) The peripheral area of Cluster View shows the mentioned keywords of highlighted documents using Gilbert curves.
  (b) The center area of Cluster View shows the topic structure of the corpus using Gosper curves.
  (c) The Document View shows a list of selected documents.
  (d) The Chatbot View provides a chatbot interface to answer user questions with the option to insert selected documents in the prompt.}
    \label{fig: sfc}
\vspace*{-0.4cm}
\end{figure*}

\vspace*{-0.4cm}
\section{Visualization of HINTs}\label{sec: visualization}
\noindent The visualization of HINTs extends the existing Space-Filling Curve (SFC) layout for large graphs proposed by Muelder et al.~\cite{muelder2008sfc}. Using the duality of hypergraphs, our visualization layout treats documents and keywords as both nodes and edges and uses two different curves to lay out the two hypergraphs. We describe our motivation for using such a layout method and our techniques to improve the readability.

\vspace*{-0.2cm}
\subsection{SFC for HyperGraph}
In the design guidelines derived by Abdelaal et al.~\cite{abdelaal2022network} in a recent network visualization evaluation study, node-link-based approaches are recommended when:
(1) tasks involve the identification of network clusters, and (2) the network is sparse.
Condition (1) is fulfilled as explained in~\autoref{sec: design_rationale}, and (2) is guaranteed by the keyword extraction process described in~\autoref{sec: keyword_extraction}. 
Therefore, we decided to use node-link-based approaches for our system.

In all the variations of node-link-based approaches for hypergraph visualization~\cite{fischer2021hypergraphsurvey}, we find the extra-node representation introduced by Ouvrard et al.~\cite{ouvrard2017hypergraph} most flexible and intuitive.
An extra-node representation improves the existing clique expansion of hypergraphs by adding extra nodes to represent hyperedges, effectively transforming the hypergraph visualization problem into a bipartite graph visualization problem.
After that, any node-link-based graph visualization method can be applied.
In our system, we use the SFC layout method to lay out the extra-node representation of the hypergraph.
The SFC layout method uses pre-computed clustering to order nodes in a sequence and then applies a space-filling curve on the node sequence to map it to a two-dimensional screen space~\cite{muelder2008sfc}.
SFC layout is known for its efficiency and aesthetics in visualizing large graphs~\cite{ma2013largegraph}.
Moreover, SFC layouts support progressive disclosure (\textit{DC2}) organically, as the layout is generated based on the clustering result (\textit{DC3}).
The preprocessing and modeling stage (\autoref{sec: methodology}) generates the document hypergraph $H_D$ and the keyword hypergraph $H_C$, each with its hierarchical clusters and visualized as two separate SFCs, as shown in~\autoref{fig: sfc}.

Specifically, we divide the layout space into two parts: the peripheral and the center area.
For the peripheral area, we concatenate four Gilbert curves~\cite{gilbert}.
A Gilbert curve is a generalized version of the Hilbert curve,
which instead of filling a squared area, can traverse any rectangular region in a way similar to the Hilbert Curve.
In~\autoref{fig: gilbert}-a, the first Gilbert curve starts from the lower left (dark blue) and ends at the lower right (dark red).
Through rotation and flipping, the start and end curve points for neighboring Gilbert curves are concatenated smoothly.
Using concatenated Gilbert curves enables us to fill a ring-like space with the efficiency and aesthetics of SFC layouts.

For the center area, we found in early prototyping that using the same curve as the peripheral region is confusing for the user, as the peripheral and center areas would become visually indistinguishable.
We decided to use a simple Gosper curve (\autoref{fig: gilbert}-c) to lay out the nodes for better aesthetics.
The resulting visualization looks similar to GosperMap~\cite{auber2013gospermap}, but we did not employ the advanced techniques proposed in GosperMap.
The interactions to support the exploration and reorganization of the dataset are the main focus of the system, which are also not limited to any specific curve.
After the curves are generated, we can apply the curves on the node sequences to generate the two-dimensional layout.
We put the document hypergraph in the center area because the documents are the main analysis targets for the user.
Consequently, the keyword hypergraph is put in the peripheral area.

\begin{figure}[t]
 \centering 
 \includegraphics[width=0.95\columnwidth]{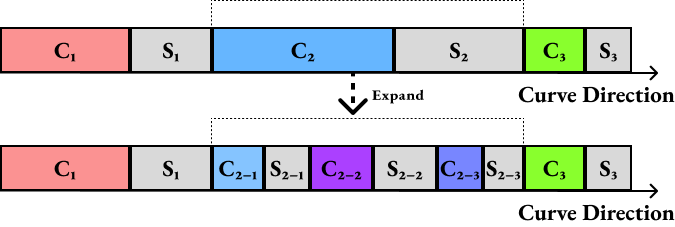}
 \caption{Spacing strategy of the SFC layouts. The space of each cluster $S_i$ is proportional to the cluster size. When $C_i$ is expanded, its sub-clusters redistribute $S_i$ to limit the changes locally. }
\label{fig: spacing}
\vspace*{-0.4cm}
\end{figure}

\vspace*{-0.4cm}
\subsection{Improving the readability}
\subsubsection{\textit{Automatic Cluster Expansion}}
In most cases, the default clustering result is not optimal for the user's targeted analysis tasks.
We identify two common problems in early prototyping: 
(1) clusters may be too big, weakening the semantic meaning that the clusters can convey.
(2) clusters may have only one sub-cluster, which makes the parent cluster redundant. 
Both problems can be mitigated by automatically expanding a cluster, i.e.breaking the cluster into sub-clusters~\autoref{fig: case_1}-d.
We employ rule-based detection to identify clusters that need to be expanded.
For a cluster $C$, we expand it if the following conditions are met:
(1) $C$ has only one sub-cluster $C_s$;
(2) $C$ has more than $n = k N$ nodes, where $k \in [0, 1]$ and $N$ is the size of the hypergraph.
Through trial and error, we find that $k=0.3$ gives the most balanced results.
\subsubsection{\textit{Spacing Strategy}}
Spacing between each node is critical for the readability and aesthetics of SFC layouts.
To highlight different clusters, we employ our spacing strategy on clusters instead of nodes.
Given a space-filling curve of a specific order, we first calculate the length of the curve $L$, the total amount of space available for the nodes, and $L - N$, the amount of space to be redistributed, where $N$ is the size of the hypergraph.
Our goal is to distribute the space between clusters to ensure the best readability.
In early prototyping, we found that distributing the space proportional to the cluster size gives the best readability as well as stability.
Specifically, we define the space of a cluster as the blank space it has behind it on the curve, which is calculated by $(L-N)\frac{N_c}{N}$, where $N_c$ is the size of the cluster (\autoref{fig: spacing}).
Another consideration when designing our spacing strategy is stability.
We want to ensure that the layout change is minimized when clusters are expanded.
Naively, when a cluster is expanded, the whole layout needs to be recalculated because now the cluster sequence is changed.
To avoid such recalculation, we design our spacing strategy in this way so that the sub-clusters can simply take over the space of their parent cluster, as shown in~\autoref{fig: spacing}.
Since the total volume of the sub-clusters is the size of the parent cluster and the space is proportional, the sub-clusters can take over exactly the space of the parent cluster without any overflow or underflow.
This ensures a local change in the layout when a cluster is expanded.

\subsubsection{\textit{Concave Hull Approximation}}
After applying the SFC layout, we use a concave hull algorithm~\cite{park2012concavehull} to generate an approximation polygon for each cluster.
The polygons are used to generate borders and calculate label positions for the clusters.
The concave hull is generated based on a cluster of points: in our case, the nodes in a cluster.
This means that the algorithm can be applied to any curve we used for the SFC layout.
This is desirable because we are using two different curves in our system, and the center area curve choice is flexible.
Using the same algorithm guarantees a unified aesthetic across curves.

The concave hull algorithm is designed for approximating points, but we need to approximate circles.
Naively we can use the center of the circles as the points for the algorithm, but this would result in a polygon that is too small and crosses the circles on the boundary.
To address this issue we use a simple trick: we add extra points to the cluster by extrapolating the original points.
For example, a Gilbert curve moves perpendicularly, so the resulting polygon would have perpendicular corners. 
We can therefore add eight extra points around each original point so that the extrapolation forms a three-by-three grid.
On the boundary of the cluster, these extra points prevent the concave hull from passing across the original points.
Since the concave hull algorithm has an $O(n\log n)$ time complexity, the performance overhead introduced by the extra points is negligible.
\begin{figure}[t]
\vspace*{-0.6cm}
    \centering
    \subfloat[\centering Four concatenated Gilbert curves. Color transitions from dark blue to dark red indicate the traverse direction.]{{\includegraphics[width=0.43\columnwidth]{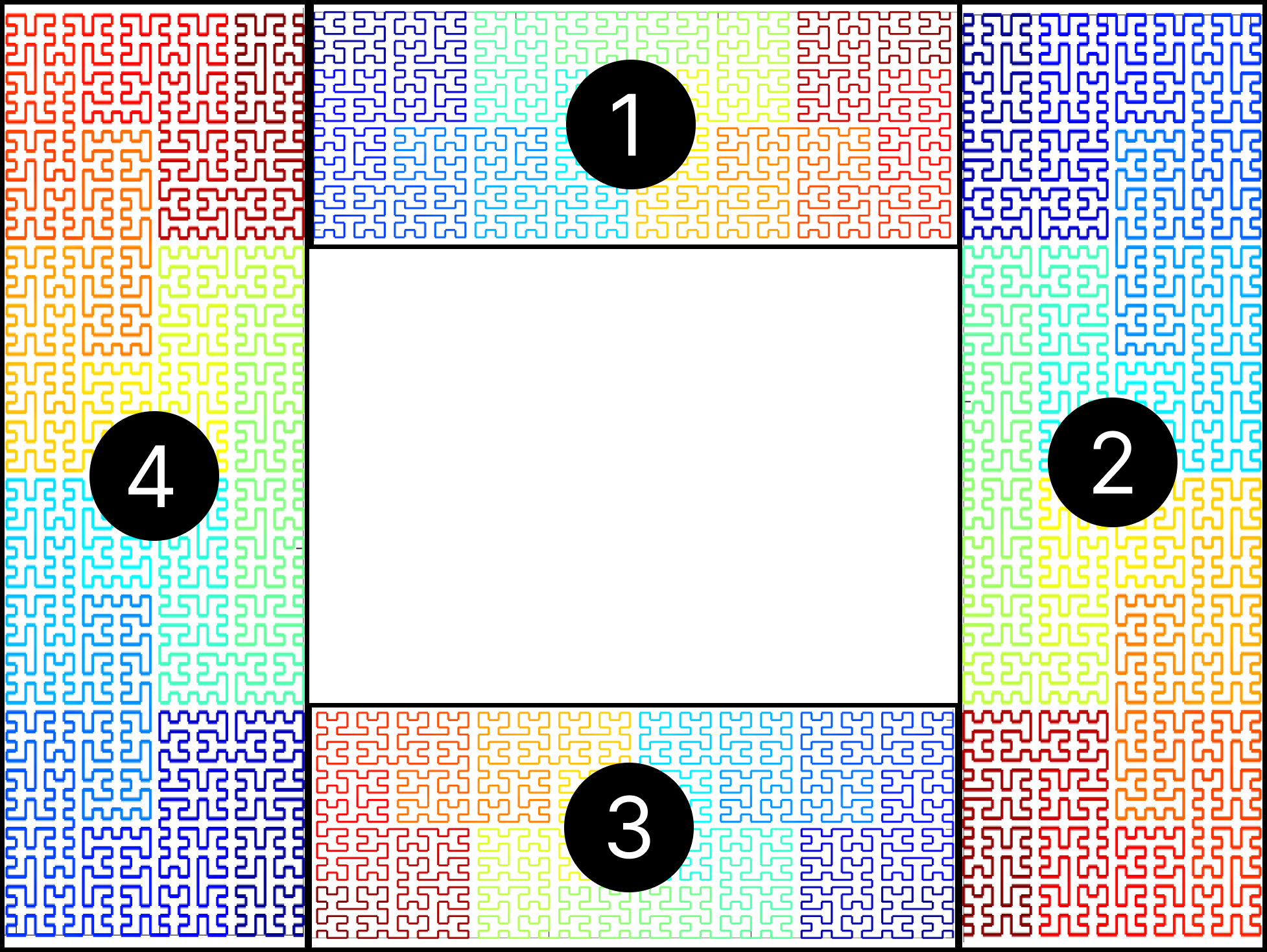} }}%
    \qquad
    \subfloat[\centering An example cluster in Gilbert curve filling layout]{{\includegraphics[width=0.43\columnwidth]{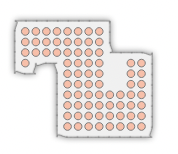} }}%
    \qquad
    \subfloat[\centering An order-4 Gosper curve]{{\includegraphics[width=0.43\columnwidth]{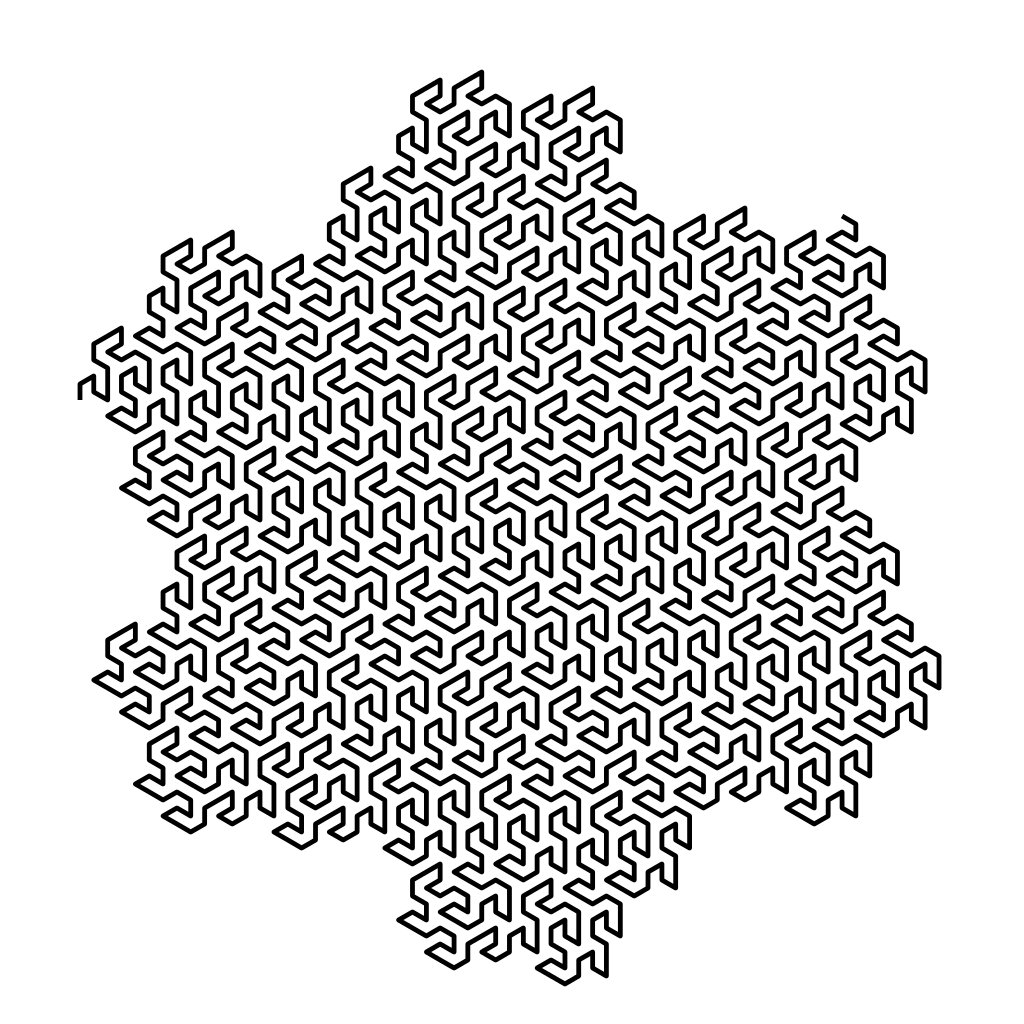} }}%
    \qquad
    \subfloat[\centering An example cluster in Gosper curve filling layout]{{\includegraphics[width=0.43\columnwidth]{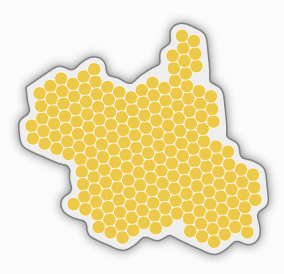} }}%
    \captionsetup{belowskip=-14pt,aboveskip=3pt}
    \caption{
        Illustrations of the space-filling curves and example clusters formed by the curves.
    }%
    \label{fig: gilbert}%
\end{figure}
\subsubsection{\textit{Borders}}
The borders are generated by applying a smoothing algorithm on the polygons.
For Gosper curves, we use the polygon as control points to generate a cubic basis spline as the border.
For Gilbert curves, we use a similar approach but with a cubic Bezier curve.
More specifically, for each pair of consecutive points, we use a smoothing factor to interpolate the control point.
This results in a sketchy style at the border corners.
Two examples are given in~\autoref{fig: gilbert}.

\subsubsection{\textit{Label Position}}
Labeling the clusters is essential for users to explore the dataset.
We use the topic assignment described in \autoref{sec: tag_assignment} to label the clusters.
For clusters, we use the centroid of the approximated polygon to position the label.
When a cluster is expanded, the sub-clusters within need to be clearly labeled as well.
Using the centroid of the sub-cluster as the label position is not a good choice because the label would cause a serious cluttering issue.
Therefore, for a sub-cluster, we first calculate its centroid $C_sub$. Then we extend the line from the parent cluster centroid $C_p$ to $C_sub$ and calculate the intersection point of the extended line and the parent border.
The intersection point is used to position the sub-cluster label, resulting in a radial layout, as shown in~\autoref{fig: case_1}-d.
The generalizability of the concave hull algorithm makes our labeling position calculation applicable to any curve used for the SFC layout.

%% file: 06_System_Design.tex
\vspace*{-0.2cm}
\section{HINTs System Design}
\noindent The HINTs system consists of three main views: Cluster View, Document View, and Analysis View.
Below, we describe the views, visualizations, and interactions, and how they assist the exploration and reorganization of a corpus.
\vspace*{-0.4cm}
\subsection{Cluster View}
Cluster View is the main view of the HINTs system (\autoref{fig: sfc}-a and -b).
It visualizes the corpus as two hypergraphs using the SFC layout.
Using the Cluster View, users can get an overview of the topic structure and keyword occurrences simultaneously (\textit{DC1}).
Interactions such as expansion, filtering, and navigation are provided to support \textit{DC2}.
Below, we discuss user interactions and the coordination with other views.
\subsubsection{Hover and click}
By default, the cluster labels are hidden to reduce clutter.
Hovering over the clusters triggers a highlight effect and shows the cluster label.
Users can select a cluster by clicking on the cluster label.
This triggers the Document View to show articles in the cluster and the mentioned keywords (\textit{T3}).
Additionally, clicking on the cluster temporarily expands the cluster to expose its sub-structure (\textit{T1}), as shown in~\autoref{fig: case_1}-b.
The sub-structure is colored in different colors while maintaining the original cluster's shape.
The labels of each sub-cluster are also shown radially.
When the user selects any cluster, the mentioned keywords are highlighted (\textit{T2}).
Under such cases, hovering over the keyword clusters will show not only the cluster label but also the highlighted keywords in a list, as shown in~\autoref{fig: sfc}-a.
Users can also select a keyword to filter the articles in the Document View (\textit{T3}).

\subsubsection{Expansion}
Users can break a cluster into smaller clusters so that a deeper level of topic structures can be explored (\textit{T1}).
This operation is necessary since the agglomerative clustering result is not always semantically optimal, and some clusters may be too vague for the user.
Upon expansion, we use an animation to show how the parent cluster is broken down into smaller pieces to maintain the visual memory.
The sub-clusters redistribute the spacing of their parent cluster proportionally to their size, so that the expansion only affects the layout locally.

\subsubsection{Filtering}
Once the users have found the target of interest,
they can click the filter button to remove irrelevant documents and keywords from the view (\textit{T3}).
The filtering functionality effectively creates a sub-hypergraph based on node selection, supporting any interactions that are available in the original hypergraph.
Note that we only support filtering on document node selection and keyword nodes are filtered accordingly.
Although filtering on keyword nodes is technically possible, we do not find the operation intuitive.
We decided to remove this feature to keep the exploration straightforward.

\subsubsection{Searching}
HINTs supports embedding-search to retrieve documents relevant to user queries (\textit{T3}).
The search functionality is implemented by first embedding the user query into the same vector space as the documents, then ranking the documents based on their relevancy to the user query using cosine similarity, thus supporting users to query in natural language.
The server returns the documents sorted by relevancy score to the front end, and the user can control the number of documents to be highlighted by a relevancy threshold. 
The highlighted documents and mentioned keywords are visually distinguished in the Cluster View (\autoref{fig: case_1}-b).

\vspace*{-0.3cm}
\subsection{Document View}
\vspace*{-0.1cm}
The Document View displays a list of selected articles (\autoref{fig: sfc}-c), either from the same cluster or returned by the search functionality (\textit{T3}).
Each document is an interactive card with the title, summary, ID, and relevancy score.
The main keywords of the document are highlighted in yellow in the summary to assist quick browsing.
If the user is under search mode, documents are sorted according to relevancy, and those above the relevancy threshold are highlighted.
Users can click any document card to add or remove it as ``query documents'', which are used in the Analysis View.

\vspace*{-0.4cm}
\subsection{Analysis View}
The Analysis View (\autoref{fig: sfc}-d) is an intelligent chatbot agent that assists users in analyzing the selected documents (\textit{T4}).
We use the OpenAI ``gpt-3.5-turbo-16k-0613'' to generate responses.
We designed the interface by enhancing a typical chatbot interface like ChatGPT with Retreival-Augmented-Generation (RAG), i.e., users can insert selected documents into the prompt by selecting document cards or adding a list of documents directly by clicking on the title of the Document View.
The chatbot then gives answers to user questions based on the selected documents.
They can also select if they want to insert the summary or the full content of the document, in case of information loss during the summarization.
The capability of LLMs to answer open-domain questions turns out to be more useful than we initially expected for the sensemaking process, as our users exhibit diverse usage patterns during the user study. More details are discussed in~\autoref{sec: user_study}.

%% file: 07_Usage_Scenario.tex
\section{Usage Scenario}\label{sec: usage_scenario}
\noindent We illustrate the use of HINTs through two usage scenarios.
We chose two different domains to demonstrate the flexibility of HINTs.
In the first usage scenario, we highlight the visualizations for interactions to support the exploration of a news article dataset.
In the second usage scenario, we focus on the usage of the Analysis View, an intelligent chatbot agent to support the analysis of selected documents.
Together, we demonstrate the functionalities that HINTs support to facilitate the sensemaking of a large corpus.
\vspace*{-0.3cm}
\subsection{Scenario 1: Collecting materials for a news story}

In this scenario, we use the \textit{AllTheNews}~\cite{allthenews} dataset to demonstrate the effectiveness of HINTs in assisting the exploration of the topics and keywords in news articles.
We randomly sampled 8192 news articles published in 2016 from the dataset for this case study.

\begin{figure}[t]
    \centering
    \includegraphics[keepaspectratio, width=\columnwidth]{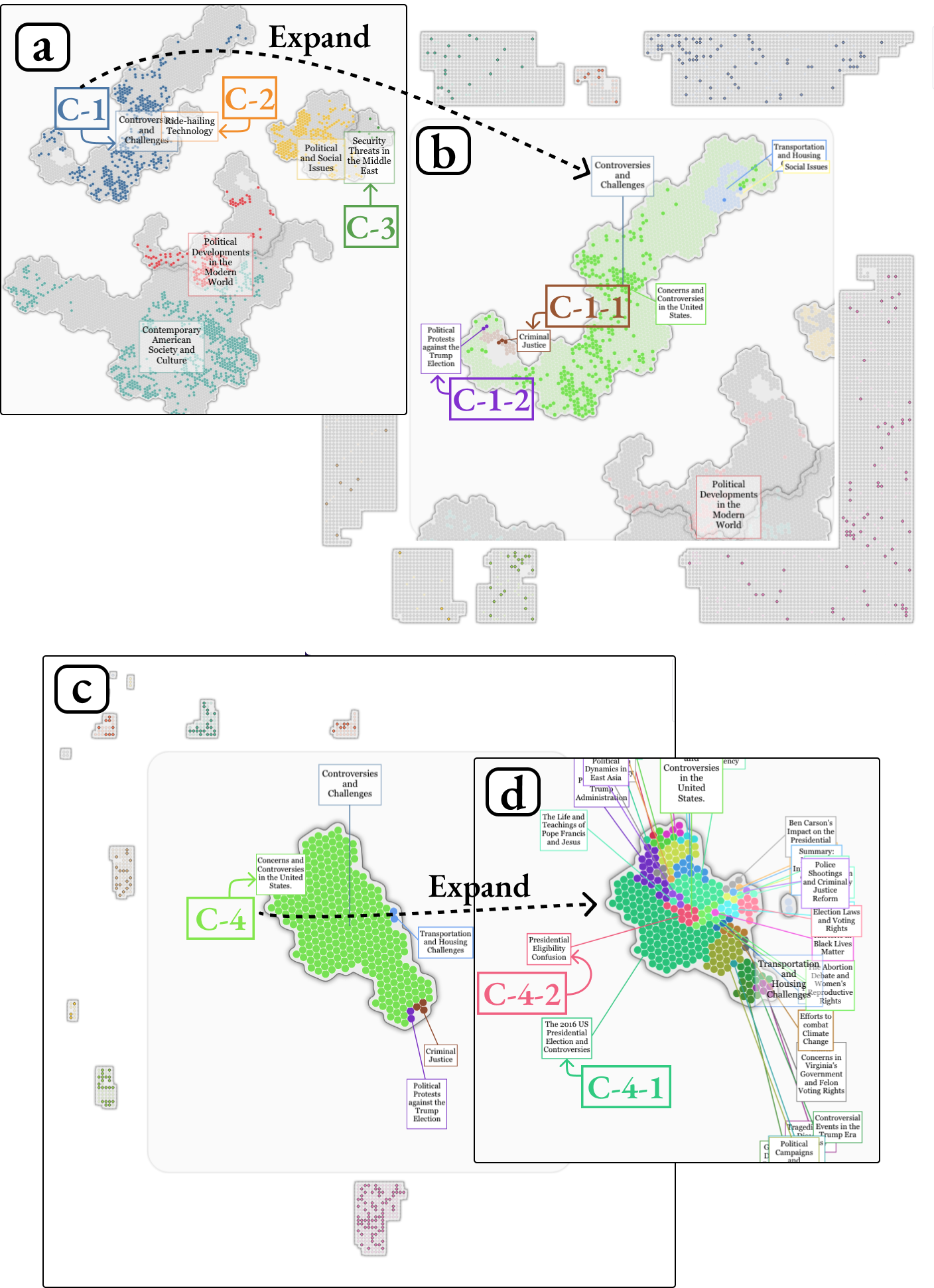}
    \caption{User interactions in usage scenario 
    (a) The search result of ``US presidential election''. Relevant articles are highlighted.
    (b) The user expands the topic ``Controversies and Challenges'', showing the sub-topics within and mentioned keywords in the peripheral area.
    (c) After clicking ``Filter Search'', irrelevant articles are removed, and the layout is regenerated.
    (d) The user expands a sub-topic ``Concerns and Controversies in the United States''.
    \vspace*{-0.5cm}
    }%
    \label{fig: case_1}%
\end{figure}

Alice is a journalist currently working on a story about the US presidential election. She wants to collect articles about the 2016 election as materials for her story.
She can only vaguely remember certain events back when the election was happening, so she wants to freshen up her memory and find real evidence to reference.
Alice begins by opening the HINTs system displaying the AllTheNews dataset and searches ``US presidential election''.
The Cluster View highlights relevant articles in each of the six topics with clear labels and colors (\autoref{fig: case_1}-a), and she can easily identify topics that are not related to the election. such as ``\textit{Ride-hailing Technology}'' (\autoref{fig: case_1}-a, C-2).
Before she filters the whole corpus with current search results, she wants to make sure that the relevancy threshold is reasonable so that she does not miss too many relevant articles.
The topic ``\textit{Security Threats in the Middle East}'' (\autoref{fig: case_1}-a, C-3) catches her attention because foreign policy towards the Middle East has always been a hot topic in the election, but the visualization suggests otherwise.
She clicks on the topic label to highlight the topic and keywords.
She hovers over each of the highlighted keyword clusters, but she does not find any keywords related to the election.
Upon quickly scanning through the titles of the highlighted articles, she confirms that the topic is not related to the election and raises the threshold until the cluster is filtered out.

After that, she finds the remaining topics too vague to make sense of. 
She first drills down the topic ``\textit{Controversies and Challenges}'' (\autoref{fig: case_1}-a, C-1) by expanding it (\autoref{fig: case_1}-b).
She finds that ``\textit{Criminal Justice}'' and ``\textit{Political Protests against the Trump election}'' (\autoref{fig: case_1}-b, C-1-1 and C-1-2) are clearly two related sub-topics.
Upon reading the summaries of the articles, she confirms the accuracy of the sub-topic labels and marks them down.

After the inspection, she is confident with the current relevancy threshold. 
She clicks the ``Filter Search'' button to reorganize the corpus (\autoref{fig: case_1}-c).
Once again, the topic ``\textit{Concerns and Controversies in the United States}''  (\autoref{fig: case_1}-c, C-4) seems to contain a lot of related articles, but she could not make sense of it solely from the topic label.
To dive deeper, she splits the cluster into separated sub-clusters (\autoref{fig: case_1}-d).
The sub-cluster ``\textit{The 2016 US Presidential Election and Controversies}'' (\autoref{fig: case_1}-d, C-4-1) catches her attention.
Upon inspecting the keywords, she becomes curious about how ``\textit{Vladimir Putin}'' is related to the election, so she clicks the keyword label to inspect the articles.
She also finds several interesting topics, such as ``\textit{Presidential Eligibility Confusion}'' (\autoref{fig: case_1}-d, C-4-2).
By browsing the highlighted keywords and reading the titles of the articles, she understands that it was about a controversy between Donald Trump and Ted Cruz during the election.
The wide range of topics brings her memory back, and she continues to click on every topic she deems relevant to the election to collect materials for her story.

This case study demonstrates that HINTs provides a highly interpretable visualization for the topic structure and keyword connections in a corpus. 
Alice can quickly understand the topics through the labels, and she can always drill down to a deeper level for more specific topics if the labels are too vague.
The keyword connections and document summaries provide her with enough information to decide the relevance of a topic to her story.
Next, we demonstrate how the Analysis View can assist users in analyzing the selected documents.

\begin{figure}[t]
    \centering
    \includegraphics[keepaspectratio, width=\columnwidth]{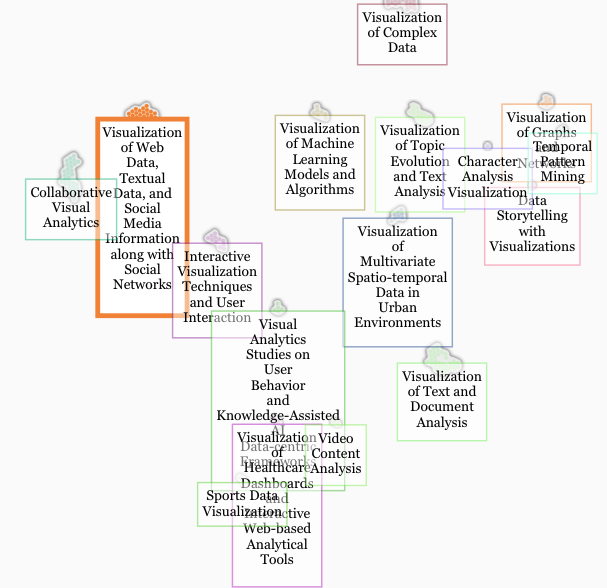}
    \caption{
    The VisPub corpus visualization was filtered by the keyword ``text analysis'' and reorganized by the user in usage scenario 2.
    After expansion and filtering, the remaining topics are at a reasonable level of granularity.
    The user is currently selecting the orange cluster ``Visualization of Web Data, Textual Data, and Social Media Information'' to prompt the Analysis View for more detail.
    }%
    \label{fig: case_2}%
\vspace*{-0.4cm}
\end{figure}
\subsection{Scenario 2: Literature Review}
In this usage scenario, we use the visualization publication dataset (VisPub)~\cite{vispub} to demonstrate how HINTs can assist users in conducting literature research.

Bob is a Ph.D. student who is currently working on literature research in text analysis in visualization for his research project.
Bob starts by loading the VisPub dataset into HINTs. 
Scanning through the topic structure, he can easily understand topics like ``Visualization of 3D Models, Volumes and Data'', ``Visualization of Medical Imaging and Blood Vessel Structures'', and ``Visualizations of Hierarchical Structures''.
Since he is only interested in text visualization, he eliminates those clusters solely based on the topic labels.
After a quick scan through all the topics presented, he proceeds to search for ``text analysis'' and applies the filter.
The search result shows that there are three related topics: ``Visualization of Complex Data and Information Spaces'', ``Visualization of Web Data, Textual Data, and Social Media Information'', and ``Visualization of Graphs and Networks''.
From there, he starts to scan through each topic to decide if it should be included in his literature research.
The mentioned keywords (author keywords) are particularly helpful in this process.
They include data types, model names, and visualization techniques, providing rich information about the topic.
He can already get a gist of the research problems in text visualization, such as ``Reasoning'', ``Topic Models'' or ``Temporal Patterns''.
He finds those topics interesting, but not all of them are directly related to his research.
By clicking on the keywords and roughly reading the abstracts, he can quickly decide whether to include it in his literature research. 
He expands and eliminates topics until he is satisfied with the reorganization result (\autoref{fig: case_2}).
After he feels confident about the overall picture, he starts to investigate the reorganized corpus in detail.

First, he finds a rather general cluster ``\textit{Visualization of Web Data, Textual Data \dots}'' 
in the result, talking about the design space of text visualization.
He selects the cluster, inserts the summaries to the prompt template, and prompts in the Analysis View: ``Given the selected documents, what should I consider when designing text visualization?''.
The chatbot responds with 10 bullet points, such as ``Amount of text'', ``Placement of text'' and ``Customizability'' while referencing the articles.
The natural language explanation with references helps him quickly make sense of the topic.
He then proceeds to ask questions in more detail while inserting the full content, such as ``What is considered a balanced ratio of text and charts?''.
In the response, a key point that the chatbot emphasizes is that users prefer charts with more textual annotations, as long as the placement of text is well-designed.

With the insertion, the chatbot provides accurate answers tailored to the inserted article, rather than replying generically.
Bob continues this process by selecting other topics he is unfamiliar with and asking the chatbot questions.
By the end, he is able to develop a clear understanding of the topics in text visualization and how researchers address them. 
We present the complete chat history in supplemental materials.

%% file: 08_User_Study.tex
\vspace*{-0.4cm}
\section{User Study}\label{sec: user_study}
\noindent The design goal of HINTs is to facilitate the sensemaking of a large corpus. Our approach combines the usage of LLMs as both a general NLP task solver and an intelligent chatbot agent, an abstraction of clustered hypergraph, and an interactive visual interface. To further evaluate the HINTs system as well as better understand user behavioral patterns and challenges during the sensemaking of a large corpus, we conducted a comparative user study. Below, we first introduce our study design, and then report the results and discuss our findings.

\vspace*{-0.4cm}
\subsection{Task Design}
Although HINTs use LLMs as both a general NLP task solver and an intelligent chatbot agent, the key goal of our work is to advocate for the integration of LLMs in the data preparation pipeline under the principle of Model Alignment. 
We thus designed a comparative, task-driven, between-subject study to highlight the benefits that visualizations designed under proper Model Alignment can bring.

During the user study, the participants are instructed to conduct a literature review on a subfield of visualization, namely explainable AI (XAI) or high-dimensional data visualization, by exploring the VisPub~\cite{vispub} dataset.
All our participants are graduate students in the visualization field. The participants are instructed to choose a subfield that they are less familiar with to reduce the effects of prior knowledge.
As a result of the literature review, the participants need to create an outline of the important topics in the selected subfield and provide a short introduction to each topic. 
The task is designed to mimic the scenario of preparing materials for a course presentation, where students are instructed to first conduct a small-scale literature review of a subfield and introduce it to other students, in which the sensemaking of the subfield is a critical step.

We divide our participants into two groups: \textit{HINTs} group, which uses the HINTs system to make sense of the subfield; and \textit{Baseline} group, which uses a basic RAG-enhanced chatbot interface that mimics the ChatGPT interface.

\vspace*{-0.4cm}
\subsection{Procedure}
The study begins by collecting demographic data of the participants, including age, gender, and years of experience with visualization research. Then, we introduce the VisPub dataset, the literature review task, and the expected outcome to the participants. After that, we introduce the interface to use, including the visual encodings, interactions, and intended usage of each component. Due to the complexity of HINTs, this process took longer for the HINTs group than the Baseline group. The participants are encouraged to play around and get familiar with the system. 

After that, the literature review task begins, with a maximum time of one hour. 
The participants are instructed to put the literature review outline in a Google doc. 
During the task, the participants are asked to follow the think-aloud protocol, so that their user experiences can be captured in audio. We also record their screen for further analysis.
After the task, the participants filled out a NASA Task Load Index (TLX)~\cite{hart1988development} to provide usability feedback. Then, we interview the participants about their satisfaction with their drafted outline, their user experience, and any relevant feedback. 

In addition to the self-reported Likert scales and interviews, to evaluate the literature review outlines, we invite experts in XAI and high-dimensional data visualization to score the outlines in terms of breadth and accuracy on a scale of 1--7 (higher is better). We report these results together in~\autoref{sec: results}.

\vspace*{-0.4cm}
\subsection{Setup and Participants}
The study was conducted fully in-person using the same setup: a 2020 Macbook Mini attached to a 27-inch display. Both user interfaces were accessed through the Firefox browser. The Google doc for drafting the literature review outline is in a Chrome browser for easy switching. All participants used the Macbook Mini's trackpad and keyboard for interaction. Before the literature review task began, we provided enough time for participants to get familiar with the trackpad and keyboard. 

A total of 12 (6 for each group) participants were recruited. 
Due to the task design, all participants are graduate students in the visualization field, with at least one year of research experience and at most 6 years of experience, including 6 females and 6 males and ages ranging from 21 to 29. 
To rule out the effect of prior knowledge during the literature review task, we asked the participants to choose a subfield and self-report their familiarity with the subfield on a scale of 1 to 7. All participants reported their familiarity to be less than 3. 

\vspace*{-0.2cm}
\section{Result and Discussion}\label{sec: results}
Below, we describe the usability evaluation results, as well as our findings on users' behavioral patterns, issues with our current approach, and potential future solutions.

\vspace*{-0.4cm}
\subsection{Usability} 
\begin{figure}[t]
    \centering
    \includegraphics[width=0.95\columnwidth]{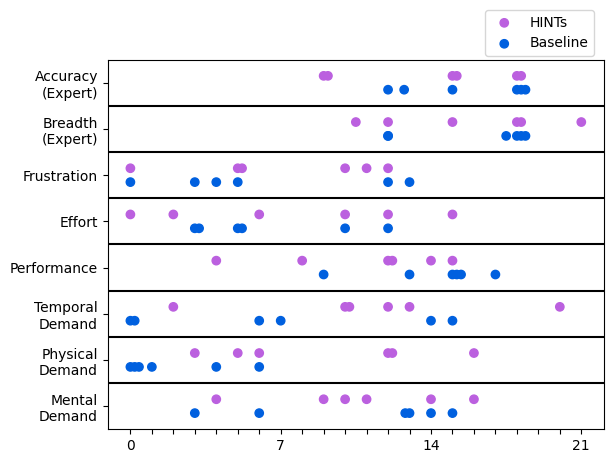}
    \caption{Result of the self-reported NASA TLX Likert scales and expert reviewed Accuracy and Breadth for HINTs group participants (purple) and Baseline group participants (blue). The x-axis is a 21-point Likert scale, divided into three sections: Low (0--7), Medium (7--14), and High (14--21). The scores show no significant difference in Accuracy, Breadth, Frustration, Effort, and Mental Demand.  HINTs group participants reported higher Physical Demand and Temporal Demand, while Baseline group participants reported higher Performance, reflecting a higher confidence in their generated outline.}
    \vspace*{-0.5cm}
    \label{fig: nasa_tlx}
\end{figure}

The self-reported results of the NASA TLX Likert scale are shown in~\autoref{fig: nasa_tlx}. The x-axis is a 21-point scale divided into three sections: Low (0--7), Medium (7--14), and High (14--21). Besides the seven scales in NASA TLX, we also added the expert review scores: \textit{Breadth} and \textit{Accuracy}. We compare the two groups by aligning them vertically. Overlapping scores are adjusted by 0.2 for visual clarity. 

\subsubsection{\textit{Comparable Usability to Baseline}} Overall, the results show that the HINTs is a highly usable and user-friendly system, as we find no significant difference between the self-reported \textit{Mental Demand}, \textit{Effort}, and \textit{Frustration}. 
Achieving comparable usability results as the Baseline group is significant as the Baseline's chatbot interface is very simple and requires minimal effort to learn and use. 
Although the HINTs Group participants reported higher scores for \textit{Physical Demand}, this is expected as the only interaction for the Baseline group participants is to type, while the HINTs system incorporates more sophisticated interactions. 
Being more physically demanding while keeping the experience effortless and enjoyable further proves the success of the interface design.
We also got no negative feedback from HINTs group participants about the visualization or interactions. 
We attribute these usability successes to a proper Model Alignment. 

\subsubsection{\textit{Over-confidence of Baseline Group}} Another interesting observation from the usability results is the comparison between self-reported Performance and expert-reviewed Breadth and Accuracy.
Though we observe no significant difference from the expert reviews, the self-reported Performance of the Baseline group is higher than that of the HINTs group.
This over-confidence can be explained by some observations we made during the user study. 
First, the Baseline group had no access to external tools to validate the response of the chatbot, so participants mostly relied on prior knowledge, or even ``trick questions'' for verification. Whenever the model gave responses that fit the participants' prior knowledge or passed these trick questions, participants gave more trust in the model and were more confident in their literature review. 
On the other hand, for the HINTs group participants, the visualizations provided visual hints on the articles that they should investigate by clustering or highlighting. However, due to the time limit, many of them were not fully investigated. As a result, participants were less confident in the completeness of their outlines.
This can be confirmed by the observation that the HINTs group participants felt more time pressure than the Baseline Group as suggested by the difference in \textit{Temporal Demand}.
In addition, when the participants were asked in the interview ``If given more time, what would you do to perfect the literature review outline?'', the Baseline Group participants mostly wanted to do further verification, while the HINTs group participants considered further investigation on certain sub-topics or techniques, suggesting awareness of the incompleteness of the outline. 

\vspace*{-0.4cm}

\subsection{Commonalities In Comparative Behavior Patterns}
Beyond the evaluation of the system, we report insights on the comparative behavior patterns to contribute to a better understanding of the sensemaking challenges with the help of LLMs. Overall, both groups of participants exhibit many commonalities in their behaviors when a powerful chatbot is at hand to help the sensemaking process, but with interactive visualizations, the participants were able to avoid certain pitfalls and made more consistent progress. We report these observations in detail in the following sections.

\subsubsection{\textit{Documents and Keywords Are Equally Vital}}
As implied by previous topic- and entity-based works, both documents and keywords are equally vital for users to make sense of a corpus. Baseline group participants, although did not have direct access to the dataset, frequently asked for example articles for a topic or a technique. The HINTs group participants used the cluster structures to find similarities and differences between different topics and then decided what hierarchy the literature review outline should have. For the keywords, both groups of participants used the chatbot heavily to generate explanations for unfamiliar keywords to help them make sense of a topic or compare different topics. Participants in HINTs group additionally used the keywords to navigate desired documents. This confirms that the interplay of documents and keywords plays a central role in corpus sensemaking. 

\subsubsection{\textit{Generative Texts Relieves Writing Burden}}
Another commonality of both groups of participants is that the chatbot is heavily used to generate summaries of topics or explanations of keywords, and then the generated texts are directly used (or after a slight edition) in the literature review outline. In real-world sensemaking scenarios, note-taking is an essential step that helps track the sensemaking progress, while previous works usually excluded it from the sensemaking process. Our observations made in the user study show that when making sense of a document or a corpus, even though reading and writing are both cognitively demanding, reading is a step that can not be avoided while writing can not. Since writing (or note-taking) usually happens after the completion of a sensemaking, users can use their gained understanding to quickly verify the generated texts. Powerful text generation tools such as LLMs can thus not only significantly lower the burden of writing, but also generate writings beyond human-level quality. This suggests new opportunities for the integration of LLMs in sensemaking.

\vspace*{-0.4cm}
\subsection{Improvements Brought By Visualizations}
Besides commonalities in the behavior patterns, we also observed three improvements in user experience which we attribute to the visual cues provided by the visualizations. Overall, visualizations provide constant visual hints for the participants to anchor on during the sensemaking. Below, we describe each in detail.
\subsubsection{\textit{Cold Start}}
Since the literature review task requires participants to investigate a subfield (XAI or high dimensional data visualization) in the visualization research, all participants started by narrowing down the dataset to find the directly related subset. For the Baseline group participants, this is done by prompting ``Give me a summary of (subfield)''. However, after this narrowing down, the Baseline group participants had difficulty finding the next action to take. 
The HINTs group participants also started by narrowing down the dataset using the search bar, but the subsequent actions for the HINTs participants were clear: to explore the highlighted clusters, as indicated by the visualization. 
This implies that the visualization can effectively provide guidance for users to direct their sensemaking. 

\subsubsection{\textit{Prompting Intention Nuances}}
Even though we used the same underlying LLM model for the Baseline and HINTs interface chatbots, participants exhibit different usages of such chatbots.
Baseline group participants frequently asked for clarification of a confusing statement or explanation of a keyword. The conversation quickly became chaotic if the model failed to provide satisfactory or consistent explanations, especially if the model mistook a ``clarification'' question from the participant as a ``skepticism'' question and overturned their previous conclusions completely.
The HINTs group participants resolved such confusion by asking simple explanation questions to the chatbot and made sense of the responses by combining the summaries and keywords in the Document View, rarely asking for further clarification. 
This implies that when a challenging cognitive process emerges (e.g., understanding an unknown keyword), people have less trust in the automatic responses of the intelligent agent and prefer to resolve the difficulty with the most direct manipulation of resources at hand; i.e., the documents and keywords. This emphasizes the necessity of interactive visualizations in sensemaking. 

\subsubsection{\textit{Losing Track of Progress}}
A common pitfall for the Baseline Group participants is losing track of their progress. 
The participants might ask for a list of subtopics to explore but unconsciously fixate on only one without realizing that there were five more subtopics. 
We attribute this phenomenon to two reasons. First, sensemaking is not linear, but hierarchical. It is our intuition to take a depth-first-search approach to systematically organize the chaotic unknowns. However, the user interface of the Baseline group presents a linear conversation between the participant and the chatbot with only texts. The visual representation quickly diverges from the participants' mental model as they go deeper into the sensemaking hierarchy. On the other hand, the graphical representation of the HINTs system is consistent with the participants' mental model, helping them keep track of their progress by reminding them of their current status and expected subsequent actions. 
Second, even though LLMs are powerful chatbots, many of their unresolved issues distract users from their sensemaking of the visualization subfield. For example, LLMs give answers that are too general in most cases. Participants have to provide extra instructions to get a more satisfactory response. Another example is the inconsistent responses. Even though the full chat history is fed to the model, the chatbot still sometimes gives contradictory conclusions or easily overturns previous conclusions upon user skepticism. To resolve such confusion, participants can only ask more questions for clarification. Such distraction rarely happened for the HINTs group participants, as the participants mainly used the chatbot for simple explanations or summaries, as mentioned earlier.

%% file: 09_Limitations_and_Future_Work.tex
\vspace*{-0.4cm}
\section{Limitations and Future Work}
\noindent The case study and user study demonstrate that HINTs effectively supports the sensemaking of a corpus.
We attribute this outcome to a good model alignment and a combined usage of LLM as a general NLP task solver and intelligent agent.
Still, there are some limitations to the current approach.
First, designing prompts for domain-specific use cases is a time-consuming trial-and-error process. There is currently a lack of effective evaluation methods for these customized extraction tasks, so the evaluation of the effectiveness of the extraction pipeline relies on human judgment. Addressing this challenge is a promising direction for future work.
We envision a combination of LLMs, computational evaluation metrics, and visualizations to assist prompt evaluation. 

Another limitation is the amount of user interactions needed to converge on a satisfying corpus organization.
Although previous study~\cite{bach2022systematic} suggests that users gain more trust and confidence with more interactions with the system, 
it would be better if the system could automatically generate a satisfying organization based on user feedback.
Our current automatic cluster expansion does not consider user intent.
We believe LLM's ability to understand user intent can be utilized to address this limitation.

Our user study participants call for deeper integration among the visualization, the documents, and the intelligent agent (chatbot). Our current implementation only supports users to inject documents as context in the prompt, but no hints are provided to show which documents are indeed used by the agent. It would be more useful if the system could show the referenced documents in the agent's response and link them back to the visualizations. 
In addition, comparative sensemaking of different clusters is also a wanted feature from the participants. Although users of HINTs could inject documents from two different clusters and ask the agent for comparisons, it is not intuitive and the comparison made by the agent is not satisfactory. Participants would like to have more control over what to compare and a graphical representation of the comparison result. 
We leave the comparative sensemaking of a large corpus as future work.

%% file: 10_Conclusion.tex
\vspace*{-0.4cm}
\section{Conclusion}
\noindent In this paper, we introduce HINTs, a visual analytics system that assists users in making sense of a corpus through exploration, reorganization, and analysis.
Previous works have shown that topic- and entity-based analysis are essential to sensemaking on a corpus, but failed to align the NLP tasks and user analysis tasks for proper Model Alignment. 
We fill in this gap by combining interactive visualization of a corpus through hypergraph and space-filling curves and state-of-the-art large language model as both a general NLP task solver and an intelligent agent.
Our evaluation proves the effectiveness of our system and the benefit of using LLMs in the data preparation stage of visualization creation to guarantee proper Model Alignment. 
We advocate for further integration of LLMs and other stages of the visualization pipeline. 
Our observations and lessons learned from the user study inform future works on the behavior patterns and challenges that users exhibit when collaborating with an intelligent chatbot agent for a sensemaking task, and the benefits of incorporating interactive visualizations. 
We especially emphasize the need for a proper evaluation of the generated textual responses in both the data preparation stage and the chatbot. 
We envision a combination of computational metrics and interactive visualizations for such evaluation.

%% file: Prompts.tex
\begin{center}
     \begin{tabular}{ | p{3.0cm} | p{3.0cm} | p{12.1cm} | }
    \hline
    \textbf{Task} & \textbf{Dataset} & \textbf{Prompt}\\
    \hline
    Summarization  & All The News & \makecell[tl]{\{\\
        \quad\textbf{Role:} `system' \\
        \quad\textbf{Content:} `You are a summarization system that summarizes events happened \\ \quad
        between the main keywords of a news article. The user will provide you with
        \\ \quad a news article to summarize.\\
        \quad Try to summarize the article with no more than three sentences. \\
        \quad Reply starts with `The article discussed \dots \\
        \}
        } \\ 
        \hline
        Keyword Extraction & All The News & \makecell[tl]{\textit{Sub-task 1: Extract event }\\ 
        \{\\ 
        \quad \textbf{Role:} `system' \\
        \quad \textbf{Content:} `You are a state-of-the-art event extraction system. Your task is to 
        \\ \quad extract only the most important event from news articles. \\
        \quad Strictly extract only one event. This event should be the most important event
        \\ \quad in the article. Also extract the trigger that indicates the occurrence of the event. \\
        \quad The events should be human-readable. Reply in this format: [\textit{event} -- \textit{trigger}] '\\ 
        \}, \\
    \{\\ 
    \quad\textbf{Role:} `user', \textbf{Content:} `This is the news article:\{\textit{sentence}\}' \\ 
    \}\\
    \hline
    \textit{Sub-task2: Extract keywords} \\ 
    \{\\
    \quad\textbf{Role:} `system' \\ 
    \quad\textbf{Content:} `You are a named entity recognition model. You will be given an \\ \quad article and the event recognized in that article by the user.  \\
    \quad The format of the input defining event in article will be: [\textit{event - category}]; \\
    \quad Extract the main characters involved in that event. The number of main 
    \\ \quad characters should be 2 or less strictly. Ignore any other characters other than the \\ \quad two main characters. Be as concise as possible.
    \\ \quad Reply with the following format: [\textit{character 1}],[\textit{character 2}] \dots'\\
    \},\\
    \{\\
    \quad \textbf{Role:} `system', \textbf{Name:} `example\_user' \\
    \quad \textbf{Content:} `Article: The article discussed the extensive history of doping in Russia,
    \\ \quad dating back to the 1983 Soviet Union's detailed instructions to inject top athletes 
    \\ \quad with anabolic steroids in order to ensure dominance at the Los Angeles Olympics.  \\
    \quad Event: [Russia's doping scandal - sports scandal]\\
    \}, \\
    \{ \\
    \quad\textbf{Role:} `system', \textbf{Name:} `example\_system', \textbf{Content:} [Russia], [Dr.Sergei]\\
    \}, \\
    \{\\
    \quad\textbf{Role:} `user', \textbf{Content: }`Article: \{\textit{sentence}\} Event: \{\textit{event}\}' \\
    \},
    } \\
    \hline
\end{tabular}
\end{center}

\clearpage
\begin{center}
 \begin{tabular}{ | p{3.0cm} | p{3.0cm} | p{12.1cm} | }
    \hline
    \textbf{Task} & \textbf{Dataset} & \textbf{Prompt}\\
    \hline
    Keyword Disambiguation  & VisPub & \makecell[tl]{\{\\
    \quad\textbf{Role:} `system'\\
    \quad\textbf{Content:} `User will provide a list of keywords from research paper abstracts.'\\ 
    \quad The input will be in the format: keyword 1, keyword 2, \dots \\
    \quad Which phrase would best describe the list of keywords? \\
    \quad The phrase should be very specific and similar to the keywords and less than 5
    \\ \quad words. If the words are too similar to each other, simply assign one of the words \\ \quad as the topic. \\
    \quad Strictly assign one topic to a list of keywords. \\
    \quad Reply in the format: [\textit{research topic}]; \\
    \},\\
    \{ \\
    \quad\textbf{Role:} `system', \textbf{Name:} `example\_user', \textbf{Content:} `[storytelling, data storytelling]' \\
    \}, \\
    \{ \\
    \quad\textbf{Role:} `system', \textbf{Name:} `example\_system', \textbf{Content:} `[data storytelling]' \\
    \}, \\
    \{ \\
    \quad\textbf{Role:} `user', \textbf{Content:} `[{\textit{keywords}}]' \\   
    \} \\
    } \\
    \hline
    \end{tabular}
\end{center}
\clearpage
\begin{center}
  \begin{tabular}{ | p{3.0cm} | p{3.0cm} | p{12.1cm} | }
    \hline
    \textbf{Task} & \textbf{Dataset} & \textbf{Prompt}\\
    \hline
    Topic Label Generation for documents  & All The News & \makecell[tl]{
    \textit{{Bottom Level:}} \\
    \{\\
    \quad\textbf{Role:} `system' \textbf{Content:} `You are a news article summarization system.  \\
    \quad The user will provide you with a set of summarized news articles, your job is to \\ \quad further summarize them into one noun phrase. \\
    \quad Use words that are already in the articles, and try to use as few words as possible.\\
    \},\\
    \{\\
    \quad\textbf{Role:} `system' \textbf{Name:} `example\_user'\\ 
    \quad\textbf{Content:} `Article 1: The article discussed an engineering company that creates 
    \\ \quad mobile research robots for the military, which released a new video showcasing
    \\ \quad the `next generation' of its humanoid Atlas robot. The footage showed the robot \\ \quad escaping and being tormented with a hockey stick, but it quickly recovered.
    \\ \quad In December 2013, Google bought Boston, the company behind the robots, \dots \\ 
    \quad Article 2: The article discussed the rise of robots and artificial intelligence (AI) \\ \quad in various industries, including Amazon's commercial delivery to a customer in \\ \quad Cambridge, England and the testing of driverless vehicles and drones. It also \\ \quad mentioned the potential job losses due to automation, but argued that the panic is \\ \quad exaggerated and that new jobs will be created. \\
    \quad Article 3: \{\textit{summary of article 3 \dots}\}\\ 
    \quad Article 4: \{\textit{summary of article 4 \dots}\}' \\
    \},\\
    \{\\
    \quad\textbf{Role:} `system', \textbf{Name:} `example\_system'\\ 
    \quad\textbf{Content:} `Robotic Advancements and Concerns'\\
    \}, \\
    \{\\
    \quad\textbf{Role:} `user', \textbf{Content:} \\ \quad `Article 1: \{\textit{article summary 1}\}, Article 2: \{\textit{article summary 2}\} \dots'\\ 
    \} \\
    \{ \\
    \quad\textbf{Role:} `system' \textbf{Content:} `You are a news article categorization system. \\
    \quad The user will provide you with a list of sub-topics of news articles and a few \\ \quad examples from the sub-topics. Your job is to further categorize the sub-topics into
    \\ \quad a single noun phrase that best summarizes all the sub-topics. Try to reuse the \\ \quad words in the examples.\\
    \} \\ \\
    \hline
    \textit{Intermediate Level:}\\
    \{ \\
    \quad\textbf{Role:} `system' \textbf{Name:} `example\_user'\\ 
    \quad\textbf{Content:} `Sub-Topics: Increasing Gun Violence in Chicago, Crime Rates and \\ \quad Policing Tactics, Misconceptions about Crime in the United States, Global Events
    \\ \quad Article 1: The article discussed the major events of 2016, including the Orlando \\ \quad nightclub shooting, Bastille Day attack in Nice, French priest assassination, and \\ \quad cafe attack in Dhaka. \dots\\
    \quad Article 2: The article discussed how serious crimes in the city increased last year \\ \quad while the number of arrests decreased \dots \\
    \}, \\
    \{\\
    \quad\textbf{Role:} `system' \textbf{Name:} `example\_system'\\ 
    \quad\textbf{Content:} `Crimes in the United States'\\
    \},\\
    \\
    } \\
    \hline
\end{tabular}
\end{center}

\clearpage
\begin{center}
  \begin{tabular}{ | p{3.0cm} | p{3.0cm} | p{12.1cm} | }
    \hline
    & VisPub & \makecell[tl]{\textit{Bottom Level:}\\
    \{\\
    \quad\textbf{Role:} `system'\\
    \quad\textbf{Content:} `You are a visualization research paper summarization system. The user 
    \\ \quad will provide you with a set of abstracts of visualization research papers. They are 
    \\ \quad manually categorized by another person, so they are discussing the same topic. 
    \\ \quad Your job is to find out what that topic is.\\
    \quad Reply with less than five words. '\\
    \},\\
    \{\\
    \quad\textbf{Role:} `system' \quad\textbf{Name:} `example\_user'\\
    \quad\textbf{Content:} `Abstract 1: Many datasets such as scientific literature collections \\ \quad contain multiple heterogeneous facets which derive implicit relations, as well as \\ \quad explicit relational references between data items \dots \\ 
    \quad Abstract 2: We present PivotPaths, an interactive visualization for exploring
    \\ \quad faceted information resources \dots \\
    \},\\
    \{\\
    \quad\textbf{Role:} `system' \quad\textbf{Name:} `example\_system'\\
    \quad\textbf{Content:}`Faceted browsing visualization'\\
    \},\\
    \{\\
    \quad\textbf{Role:} `user'\\
    \quad\textbf{Content:} `Abstract 1: \{\textit{abstract 1}\}, Abstract 2: \{\textit{abstract 2}\} \dots'\\
    \}\\
    \\
    \hline
    \textit{Intermediate Level:}\\
    \{\\
    \quad\textbf{Role:} `system'\\
    \quad\textbf{Content:} `You are a visualization research paper summarization system. \\
    \quad You generate topics for a set of visualization research papers.\\
    \quad The user will provide you with a list of sub-topics and a few example abstracts
    \\ \quad from the sub-topics. Your job is to further categorize the sub-topics into a single 
    \\ \quad noun phrase that best summarizes all the sub-topics. Try to reuse the words in the \\ \quad sub-topics. Reply with a single noun phrase without any line breaks. Be concise.\\
    \},\\
    \{\\
    \quad\textbf{Role:} `system' \quad\textbf{Name:} `example\_user'\\
    \quad\textbf{Content:} `Sub-Topics: Underwater 3D scene reconstruction from acoustic imag-\\
    \quad ing sonar data, Visualization of Sound Propagation in Room Acoustics, Under-\\
    \quad water seabed visualization;\\
    \quad Abstract 1: The development of a high speed multi-frequency continuous scan
    \\ \quad sonar at Sonar Research Development Ltd has resulted in the acquisition of \\ \quad extremely accurate, high resolution\\
    \quad bathymetric data. \dots\\
    \quad Abstract 2:  \dots '\\
    \},\\
    \{\\
    \quad\textbf{Role:} `system' \quad\textbf{Name:} `example\_system'\\
    \quad\textbf{Content:} `Visualization of Underwater Acoustic Scenes and Sound Propagation'\\
    \},\\
    \{\\
    \quad\textbf{Role:} `user'\\
    \quad\textbf{Content:} \\ \quad `Sub-Topics: \{\textit{sub topics}\}, Abstract 1: \{\textit{abstract 1}\}, Abstract 2: \{\textit{abstract 2}\}, \dots'\\
    \}\\
    }\\
    \hline
\end{tabular}
\end{center}

\clearpage
\begin{center}
  \begin{tabular}{ | p{3.0cm} | p{3.0cm} | p{12.1cm} | }
    \hline
    \textbf{Task} & \textbf{Dataset} & \textbf{Prompt}\\
    \hline
    Topic Label Generation for keywords & All The News & \makecell[tl]{
    \{\\
    \quad\textbf{Role:} `system'\\
    \quad\textbf{Content:} `You are an entity summarization system.\\
    \quad The user will provide you with a list of entities, they can be people, places, or \\ \quad things. The user wants to get a gist of what entities are in the list.\\
    \quad First, split the entities into different categories.\\
    \quad Then, assign each category a human-readable name.\\
    \quad If entities in a category are all related to a specific entity, use that entity as the 
    \\ \quad category.  Limit the number of categories to be less than 5 by keeping only the \\ \quad important categories. \\
    \quad Reply with the following format: Category 1, Category 2, Category 3, \dots \\
    \quad Do not reply more than 5 categories.'\\
    \},\\
    \{\\
    \quad\textbf{Role:} `system', \textbf{Name:} `example\_user\_1'\\ 
    \quad\textbf{Content:} \\ \quad `Entities: Refugee, Refugee camp, Internally displaced person, Asylum seeker'\\
    \}, \\
    \{\\
    \quad\textbf{Role:} `system', \textbf{Name:} `example\_system\_1', \textbf{Content:} `Refugee'\\
    \},\\
    \{\\
    \quad\textbf{Role:} `system', \textbf{Name:} `example\_user\_2'\\ 
    \quad\textbf{Content:} \\ \quad `Entities: JetBlues, American Airlines, Delta Airlines, Southwest Airlines, United\\
    \quad Airlines, Alaksa Airlines'\\
    \}, \\
    \{\\
    \quad\textbf{Role:} `system', \textbf{Name:} `example\_system\_2', \textbf{Content:} `US airline companies'\\
    \},\\
    \{\\
    \quad\textbf{Role:} `user', \textbf{Content:} `Entities: \{\textit{Entity List}\}'\\
    \},\\
    } \\
    \cline{2-3}
     & VisPub & \makecell[tl]{
    \{\\
    \quad\textbf{Role:} `system'\\
    \quad\textbf{Content:} `You are a visualization research paper keyword summarization system.\\
    \quad The user will provide you with a list of keywords, they are terminologies in \\ \quad visualization research papers. The user wants to get a gist of what keywords 
    \\ \quad are in the list, but the list is too long.\\
    \quad Pick out only a few keywords that best represents the list. 
    \\ \quad Avoid picking out overlapping keywords. \\
    \quad Limit the number of picked keywords to be less than 5 by keeping only the
    \\ \quad important ones. \\
    \quad Reply with the following format: Keyword\_1, Keyword\_2, Keyword\_3, \dots \\
    \},\\
    \{\\
    \quad\textbf{Role: } `user', \textbf{Content:} `Keywords: \{\textit{keyword list}\}' \\
    \}\\
    }\\
    \hline
\end{tabular}
\end{center}
\clearpage
\section{Chat Responses for usage scenario 2}
\quad \textit{Q: } Given the selected documents, what should I consider when designing text visualization?

\textit{A: }
When designing text visualization, here are some key considerations based on the selected articles:\\
When designing text visualization, you should consider the following aspects based on the selected articles: \\
1. Amount of text: Determine the appropriate amount of text to include in the visualization. The articles suggest that adding text annotations to visualizations can enhance understanding and should not be penalized. Participants preferred charts with more textual annotations over charts with fewer annotations or text alone. \\
2. Semantic content: Consider the type of information conveyed by the text. Text that describes statistical or relational components of a chart leads to more takeaways referring to statistics or relational comparisons. \\
3. Placement of text: Text should be placed strategically within the visualization. Different kinds of information are best placed in the title or closer to the data, depending on the semantic level. \\
4. Integration of text and charts: Aim for a tight coupling between text and charts to support the synthesis of information. Investigate ways to establish interactive references between text and charts, allowing users to construct connections effortlessly. \\
5. Real-time analysis: If dealing with real-time data, consider the scalability and adaptiveness of the visualization. Develop interactive systems that enable visual analysis of streaming social media data or dynamically cluster posts related to relevant topics. \\
6. Support for specific tasks: Tailor the visualization to specific tasks such as clinical chart review, argument structure analysis, or anomaly detection in social media. Incorporate insights from domain experts and conduct iterative design processes to ensure usability. \\
7. Enhancement of comprehension: Explore techniques that facilitate enhanced comprehension of complex documents by coupling text content with data visualizations. Generate on-demand visualizations based on the reader's focus within the document and evaluate their effectiveness against traditional reading experiences. \\
8. Customizability and integration with existing workflows: Provide tools that are customizable to the needs of users, such as healthcare providers and physicians, while preserving and extensively linking to the original text. Avoid obscuring or substituting the original clinical text that contains qualitative and rhetorical flavors. \\
9. Facilitating analysis of complex textual differences: Develop visual analytics environments that support the alignment of complex textual differences, prevailing in medieval text traditions or other literature domains. Provide interactive visualizations for investigating repetition, variance, and other textual hierarchies. \\
10. Evaluation and validation: Perform user studies, expert reviews, and evaluations to validate the effectiveness and utility of the visualization techniques. Collect feedback from users and iterate designs based on their needs and preferences. Consider these guidelines when designing text visualizations to ensure effective communication and comprehension of information.\\

\textit{Q: } Given the selected documents, what is considered a balanced ratio of text and charts?

\textit{A: }
Based on the selected article, it is suggested that a balanced ratio of text and charts can vary depending on the context and purpose of the visualization. 
The article mentions that heavily annotated charts were not penalized and participants preferred charts with more textual annotations. 
However, it also emphasizes the importance of considering the qualitative properties and placement of the text within the chart. 
Overall, it is recommended to use a combination of text and charts that provides additional context, guidance, and relevant information to the readers without overwhelming or distracting them from the visualizations.

\section{Agglomerative Algorithm}
\begin{algorithm}\label{alg: clustering}
    \caption{Agglomerative Clustering }\label{alg:cap}
    \hspace*{\algorithmicindent} \textit{Input}: $H=(V, E)$, $Vec=\{vec_i |\, i \in V\}$ \\
    \hspace*{\algorithmicindent} \textit{Output}: Sequence of partitions $\mathbf{P}=P_1, P_2, \dots P_k$ \\
    \begin{algorithmic}[1]
    \While{$|V| > 1$}
        \State$P_k = init\_partition(V)$ \Comment{Initialize each node as a cluster}
        \State$\mathbf{S_s} = cosine\_similarity(V\times V, Vec)$ 
        \State$\mathbf{S_c} = wTO(V\times V, E)$ \
        \For{$i \in V$}
            \State$j = most\_similar\_node(i, S_s, S_c)$ 
            \State$P_k = merge\_clusters(i, j)$ \Comment{Merge the two clusters}
        \EndFor
        \State$H^\prime=(V^\prime, E^\prime) = construct\_hypergraph(P_k)$ \Comment{clusters are the new nodes}
        \State$Vec^\prime = centroid\_similarity(V^\prime)$
        \State$V=V^\prime, E=E^\prime, Vec=Vec^\prime$ \Comment{Update for next iteration}
    \EndWhile
    \end{algorithmic}
\end{algorithm}

%% file: main.bbl
\begin{thebibliography}{10}

\bibitem{allthenews}
{All The News}.
\newblock \url{https://components.one/datasets/all-the-news-2-news-articles-dataset}.

\bibitem{abdelaal2022network}
M.~Abdelaal, N.~D. Schiele, K.~Angerbauer, K.~Kurzhals, M.~Sedlmair, and D.~Weiskopf.
\newblock Comparative evaluation of bipartite, node-link, and matrix-based network representations.
\newblock {\em IEEE Transactions on Visualization and Computer Graphics}, 29(1):896--906, 2022.

\bibitem{alexander2014serendip}
E.~Alexander, J.~Kohlmann, R.~Valenza, M.~Witmore, and M.~Gleicher.
\newblock Serendip: Topic model-driven visual exploration of text corpora.
\newblock In {\em 2014 IEEE Conference on Visual Analytics Science and Technology (VAST)}, pp. 173--182, 2014. \href{https://doi.org/10.1109/VAST.2014.7042493}
{doi: {{%
10\hspace{.1pt}\discretionary{.}{%
}{.}\hspace{.4pt}1109\discretionary{/}{%
}{/}VAST\hspace{.1pt}\discretionary{.}{%
}{.}\hspace{.4pt}2014\hspace{.1pt}\discretionary{.}{%
}{.}\hspace{.4pt}7042493}}}


\bibitem{atzberger2023evaluatetopicmodel}
D.~Atzberger, T.~Cech, W.~Scheibel, M.~Trapp, R.~Richter, J.~D{\"o}llner, and T.~Schreck.
\newblock Large-scale evaluation of topic models and dimensionality reduction methods for 2d text spatialization.
\newblock {\em arXiv preprint arXiv:2307.11770}, 2023.

\bibitem{auber2013gospermap}
D.~Auber, C.~Huet, A.~Lambert, B.~Renoust, A.~Sallaberry, and A.~Saulnier.
\newblock Gospermap: Using a gosper curve for laying out hierarchical data.
\newblock {\em IEEE transactions on visualization and computer graphics}, 19(11):1820--1832, 2013.

\bibitem{ayoola2022refined}
T.~Ayoola, S.~Tyagi, J.~Fisher, C.~Christodoulopoulos, and A.~Pierleoni.
\newblock {R}e{F}in{ED}: An efficient zero-shot-capable approach to end-to-end entity linking.
\newblock In {\em NAACL}, 2022.

\bibitem{bach2022systematic}
T.~A. Bach, A.~Khan, H.~Hallock, G.~Beltr{\~a}o, and S.~Sousa.
\newblock A systematic literature review of user trust in ai-enabled systems: An hci perspective.
\newblock {\em International Journal of Human--Computer Interaction}, pp. 1--16, 2022.

\bibitem{bang2023multitask}
Y.~Bang, S.~Cahyawijaya, N.~Lee, W.~Dai, D.~Su, B.~Wilie, H.~Lovenia, Z.~Ji, T.~Yu, W.~Chung, et~al.
\newblock A multitask, multilingual, multimodal evaluation of chatgpt on reasoning, hallucination, and interactivity.
\newblock {\em arXiv preprint arXiv:2302.04023}, 2023.

\bibitem{beasley2021through}
Z.~J. Beasley, A.~Friedman, and P.~Rosen.
\newblock Through the looking glass: insights into visualization pedagogy through sentiment analysis of peer review text.
\newblock {\em IEEE Computer Graphics and Applications}, 41(6):59--70, 2021.

\bibitem{kaminski2021hgraphcommunity}
P.~P. . F.~T. Bogumił~Kamiński.
\newblock Community detection algorithm using hypergraph modularity.
\newblock In {\em Complex Networks \& Their Applications IX: Volume 1, Proceedings of the Ninth International Conference on Complex Networks and Their Applications COMPLEX NETWORKS 2020}, pp. 152--163. Springer, 2021.

\bibitem{brown2020language}
T.~Brown, B.~Mann, N.~Ryder, M.~Subbiah, J.~D. Kaplan, P.~Dhariwal, A.~Neelakantan, P.~Shyam, G.~Sastry, A.~Askell, et~al.
\newblock Language models are few-shot learners.
\newblock {\em Advances in neural information processing systems}, 33:1877--1901, 2020.

\bibitem{caillou2020cartolabe}
P.~Caillou, J.~Renault, J.-D. Fekete, A.-C. Letournel, and M.~Sebag.
\newblock Cartolabe: A web-based scalable visualization of large document collections.
\newblock {\em IEEE Computer Graphics and Applications}, 41(2):76--88, 2020.

\bibitem{cao2011solarmap}
N.~Cao, D.~Gotz, J.~Sun, Y.-R. Lin, and H.~Qu.
\newblock Solarmap: Multifaceted visual analytics for topic exploration.
\newblock In {\em 2011 IEEE 11th International Conference on Data Mining}, pp. 101--110. IEEE, 2011.

\bibitem{cao2010facetatlas}
N.~Cao, J.~Sun, Y.-R. Lin, D.~Gotz, S.~Liu, and H.~Qu.
\newblock Facetatlas: Multifaceted visualization for rich text corpora.
\newblock {\em IEEE transactions on visualization and computer graphics}, 16(6):1172--1181, 2010.

\bibitem{cho2016vairoma}
I.~Cho, W.~Dou, D.~X. Wang, E.~Sauda, and W.~Ribarsky.
\newblock Vairoma: A visual analytics system for making sense of places, times, and events in roman history.
\newblock {\em IEEE Transactions on Visualization and Computer Graphics}, 22(1):210--219, 2016. \href{https://doi.org/10.1109/TVCG.2015.2467971}
{doi: {{%
10\hspace{.1pt}\discretionary{.}{%
}{.}\hspace{.4pt}1109\discretionary{/}{%
}{/}TVCG\hspace{.1pt}\discretionary{.}{%
}{.}\hspace{.4pt}2015\hspace{.1pt}\discretionary{.}{%
}{.}\hspace{.4pt}2467971}}}


\bibitem{choo2013utopian}
J.~Choo, C.~Lee, C.~K. Reddy, and H.~Park.
\newblock Utopian: User-driven topic modeling based on interactive nonnegative matrix factorization.
\newblock {\em IEEE Transactions on Visualization and Computer Graphics}, 19(12):1992--2001, 2013. \href{https://doi.org/10.1109/TVCG.2013.212}
{doi: {{%
10\hspace{.1pt}\discretionary{.}{%
}{.}\hspace{.4pt}1109\discretionary{/}{%
}{/}TVCG\hspace{.1pt}\discretionary{.}{%
}{.}\hspace{.4pt}2013\hspace{.1pt}\discretionary{.}{%
}{.}\hspace{.4pt}212}}}


\bibitem{chuang2012termite}
J.~Chuang, C.~D. Manning, and J.~Heer.
\newblock Termite: Visualization techniques for assessing textual topic models.
\newblock In {\em Proceedings of the international working conference on advanced visual interfaces}, pp. 74--77, 2012.

\bibitem{chuang2012interpretation}
J.~Chuang, D.~Ramage, C.~Manning, and J.~Heer.
\newblock Interpretation and trust: Designing model-driven visualizations for text analysis.
\newblock In {\em Proceedings of the SIGCHI conference on human factors in computing systems}, pp. 443--452, 2012.

\bibitem{citraro2020eva}
S.~Citraro and G.~Rossetti.
\newblock Eva: Attribute-aware network segmentation.
\newblock In {\em Complex Networks and Their Applications VIII: Volume 1 Proceedings of the Eighth International Conference on Complex Networks and Their Applications COMPLEX NETWORKS 2019 8}, pp. 141--151. Springer, 2020.

\bibitem{combe2015louvain}
D.~Combe, C.~Largeron, M.~G{\'e}ry, and E.~Egyed-Zsigmond.
\newblock I-louvain: An attributed graph clustering method.
\newblock In {\em Advances in Intelligent Data Analysis XIV: 14th International Symposium, IDA 2015, Saint Etienne. France, October 22-24, 2015. Proceedings 14}, pp. 181--192. Springer, 2015.

\bibitem{devlin2018bert}
J.~Devlin, M.-W. Chang, K.~Lee, and K.~Toutanova.
\newblock Bert: Pre-training of deep bidirectional transformers for language understanding.
\newblock {\em arXiv preprint arXiv:1810.04805}, 2018.

\bibitem{dou2013hierarchicaltopics}
W.~Dou, L.~Yu, X.~Wang, Z.~Ma, and W.~Ribarsky.
\newblock Hierarchicaltopics: Visually exploring large text collections using topic hierarchies.
\newblock {\em IEEE Transactions on Visualization and Computer Graphics}, 19(12):2002--2011, 2013.

\bibitem{fischer2021hypergraphsurvey}
M.~T. Fischer, A.~Frings, D.~A. Keim, and D.~Seebacher.
\newblock Towards a survey on static and dynamic hypergraph visualizations.
\newblock In {\em 2021 IEEE visualization conference (VIS)}, pp. 81--85. IEEE, 2021.

\bibitem{gilardi2023chatgpt}
F.~Gilardi, M.~Alizadeh, and M.~Kubli.
\newblock Chatgpt outperforms crowd-workers for text-annotation tasks.
\newblock {\em arXiv preprint arXiv:2303.15056}, 2023.

\bibitem{gysi2018wto}
D.~M. Gysi, A.~Voigt, T.~d.~M. Fragoso, E.~Almaas, and K.~Nowick.
\newblock wto: an r package for computing weighted topological overlap and a consensus network with integrated visualization tool.
\newblock {\em BMC bioinformatics}, 19(1):1--16, 2018.

\bibitem{han2022hisva}
D.~Han, G.~Parsad, H.~Kim, J.~Shim, O.-S. Kwon, K.~A. Son, J.~Lee, I.~Cho, and S.~Ko.
\newblock Hisva: A visual analytics system for studying history.
\newblock {\em IEEE Transactions on Visualization and Computer Graphics}, 28(12):4344--4359, 2022. \href{https://doi.org/10.1109/TVCG.2021.3086414}
{doi: {{%
10\hspace{.1pt}\discretionary{.}{%
}{.}\hspace{.4pt}1109\discretionary{/}{%
}{/}TVCG\hspace{.1pt}\discretionary{.}{%
}{.}\hspace{.4pt}2021\hspace{.1pt}\discretionary{.}{%
}{.}\hspace{.4pt}3086414}}}


\bibitem{hart1988development}
S.~G. Hart and L.~E. Staveland.
\newblock Development of nasa-tlx (task load index): Results of empirical and theoretical research.
\newblock In {\em Advances in psychology}, vol.~52, pp. 139--183. Elsevier, 1988.

\bibitem{vispub}
P.~Isenberg, F.~Heimerl, S.~Koch, T.~Isenberg, P.~Xu, C.~Stolper, M.~Sedlmair, J.~Chen, T.~M{\"o}ller, and J.~Stasko.
\newblock vispubdata.org: A metadata collection about {IEEE} visualization ({VIS}) publications.
\newblock {\em IEEE Transactions on Visualization and Computer Graphics}, 23(9):2199--2206, Sept. 2017. \href{https://doi.org/10.1109/TVCG.2016.2615308}
{doi: {{%
10\hspace{.1pt}\discretionary{.}{%
}{.}\hspace{.4pt}1109\discretionary{/}{%
}{/}TVCG\hspace{.1pt}\discretionary{.}{%
}{.}\hspace{.4pt}2016\hspace{.1pt}\discretionary{.}{%
}{.}\hspace{.4pt}2615308}}}


\bibitem{izacard2022unsupervised}
G.~Izacard, M.~Caron, L.~Hosseini, S.~Riedel, P.~Bojanowski, A.~Joulin, and E.~Grave.
\newblock Unsupervised dense information retrieval with contrastive learning.
\newblock {\em Transactions on Machine Learning Research}, 2022.

\bibitem{karpukhin-etal-2020-dense}
V.~Karpukhin, B.~Oguz, S.~Min, P.~Lewis, L.~Wu, S.~Edunov, D.~Chen, and W.-t. Yih.
\newblock Dense passage retrieval for open-domain question answering.
\newblock In {\em Proceedings of the 2020 Conference on Empirical Methods in Natural Language Processing (EMNLP)}, pp. 6769--6781. Association for Computational Linguistics, Online, Nov. 2020. \href{https://doi.org/10.18653/v1/2020.emnlp-main.550}
{doi: {{%
10\hspace{.1pt}\discretionary{.}{%
}{.}\hspace{.4pt}18653\discretionary{/}{%
}{/}v1\discretionary{/}{%
}{/}2020\hspace{.1pt}\discretionary{.}{%
}{.}\hspace{.4pt}emnlp\discretionary{%
}{-}{-}main\hspace{.1pt}\discretionary{.}{%
}{.}\hspace{.4pt}550}}}


\bibitem{kumar2020new}
T.~Kumar, S.~Vaidyanathan, H.~Ananthapadmanabhan, S.~Parthasarathy, and B.~Ravindran.
\newblock A new measure of modularity in hypergraphs: Theoretical insights and implications for effective clustering.
\newblock In {\em Complex Networks and Their Applications VIII: Volume 1 Proceedings of the Eighth International Conference on Complex Networks and Their Applications COMPLEX NETWORKS 2019 8}, pp. 286--297. Springer, 2020.

\bibitem{lee2012ivisclustering}
H.~Lee, J.~Kihm, J.~Choo, J.~Stasko, and H.~Park.
\newblock ivisclustering: An interactive visual document clustering via topic modeling.
\newblock In {\em Computer graphics forum}, vol.~31, pp. 1155--1164. Wiley Online Library, 2012.

\bibitem{lee2017humantopicmodel}
T.~Y. Lee, A.~Smith, K.~Seppi, N.~Elmqvist, J.~Boyd-Graber, and L.~Findlater.
\newblock The human touch: How non-expert users perceive, interpret, and fix topic models.
\newblock {\em International Journal of Human-Computer Studies}, 105:28--42, 2017.

\bibitem{li2023evaluateChatgpt}
B.~Li, G.~Fang, Y.~Yang, Q.~Wang, W.~Ye, W.~Zhao, and S.~Zhang.
\newblock Evaluating chatgpt's information extraction capabilities: An assessment of performance, explainability, calibration, and faithfulness.
\newblock {\em arXiv preprint arXiv:2304.11633}, 2023.

\bibitem{ma2013largegraph}
K.-L. Ma and C.~W. Muelder.
\newblock Large-scale graph visualization and analytics.
\newblock {\em Computer}, 46(7):39--46, 2013. \href{https://doi.org/10.1109/MC.2013.242}
{doi: {{%
10\hspace{.1pt}\discretionary{.}{%
}{.}\hspace{.4pt}1109\discretionary{/}{%
}{/}MC\hspace{.1pt}\discretionary{.}{%
}{.}\hspace{.4pt}2013\hspace{.1pt}\discretionary{.}{%
}{.}\hspace{.4pt}242}}}


\bibitem{maddigan2023chat2vis}
P.~Maddigan and T.~Susnjak.
\newblock Chat2vis: Generating data visualisations via natural language using chatgpt, codex and gpt-3 large language models.
\newblock {\em IEEE Access}, 2023.

\bibitem{muelder2008sfc}
C.~Muelder and K.-L. Ma.
\newblock Rapid graph layout using space filling curves.
\newblock {\em IEEE Transactions on Visualization and Computer Graphics}, 14(6):1301--1308, 2008.

\bibitem{narechania2021vitality}
A.~Narechania, A.~Karduni, R.~Wesslen, and E.~Wall.
\newblock Vitality: Promoting serendipitous discovery of academic literature with transformers \& visual analytics.
\newblock {\em IEEE Transactions on Visualization and Computer Graphics}, 28(1):486--496, 2021.

\bibitem{narechania2022vitality}
A.~Narechania, A.~Karduni, R.~Wesslen, and E.~Wall.
\newblock vitality: Promoting serendipitous discovery of academic literature.
\newblock 2022.

\bibitem{nasar2021named}
Z.~Nasar, S.~W. Jaffry, and M.~K. Malik.
\newblock Named entity recognition and relation extraction: State-of-the-art.
\newblock {\em ACM Computing Surveys (CSUR)}, 54(1):1--39, 2021.

\bibitem{oelke2014topiccoin}
D.~Oelke, H.~Strobelt, C.~Rohrdantz, I.~Gurevych, and O.~Deussen.
\newblock Comparative exploration of document collections: a visual analytics approach.
\newblock In {\em Computer Graphics Forum}, vol.~33, pp. 201--210. Wiley Online Library, 2014.

\bibitem{ouvrard2017hypergraph}
X.~Ouvrard, J.~L. Goff, and S.~Marchand{-}Maillet.
\newblock Networks of collaborations: Hypergraph modeling and visualisation.
\newblock {\em CoRR}, abs/1707.00115, 2017.

\bibitem{park2018conceptvector}
D.~Park, S.~Kim, J.~Lee, J.~Choo, N.~Diakopoulos, and N.~Elmqvist.
\newblock Conceptvector: Text visual analytics via interactive lexicon building using word embedding.
\newblock {\em IEEE Transactions on Visualization and Computer Graphics}, 24(1):361--370, 2018. \href{https://doi.org/10.1109/TVCG.2017.2744478}
{doi: {{%
10\hspace{.1pt}\discretionary{.}{%
}{.}\hspace{.4pt}1109\discretionary{/}{%
}{/}TVCG\hspace{.1pt}\discretionary{.}{%
}{.}\hspace{.4pt}2017\hspace{.1pt}\discretionary{.}{%
}{.}\hspace{.4pt}2744478}}}


\bibitem{park2012concavehull}
J.-S. Park and S.-J. Oh.
\newblock A new concave hull algorithm and concaveness measure for n-dimensional datasets.
\newblock {\em Journal of Information science and engineering}, 28(3):587--600, 2012.

\bibitem{qiu2022docflow}
R.~Qiu, Y.~Tu, Y.-S. Wang, P.-Y. Yen, and H.-W. Shen.
\newblock Docflow: A visual analytics system for question-based document retrieval and categorization.
\newblock {\em IEEE Transactions on Visualization and Computer Graphics}, 2022.

\bibitem{raval2023explainandtrust}
S.~Raval, C.~Wang, F.~Vi\'egas, and M.~Wattenberg.
\newblock Explain and trust: An interactive machine learning framework for exploring text embeddings.
\newblock {\em IEEE Transactions on Visualization and Computer Graphics}, 2023.

\bibitem{shen2014entity}
W.~Shen, J.~Wang, and J.~Han.
\newblock Entity linking with a knowledge base: Issues, techniques, and solutions.
\newblock {\em IEEE Transactions on Knowledge and Data Engineering}, 27(2):443--460, 2014.

\bibitem{sherkat2018interactive}
E.~Sherkat, S.~Nourashrafeddin, E.~E. Milios, and R.~Minghim.
\newblock Interactive document clustering revisited: A visual analytics approach.
\newblock In {\em 23rd International Conference on Intelligent User Interfaces}, pp. 281--292, 2018.

\bibitem{Stasko2007jigasw}
J.~Stasko, C.~Gorg, Z.~Liu, and K.~Singhal.
\newblock Jigsaw: Supporting investigative analysis through interactive visualization.
\newblock In {\em 2007 IEEE Symposium on Visual Analytics Science and Technology}, pp. 131--138, 2007. \href{https://doi.org/10.1109/VAST.2007.4389006}
{doi: {{%
10\hspace{.1pt}\discretionary{.}{%
}{.}\hspace{.4pt}1109\discretionary{/}{%
}{/}VAST\hspace{.1pt}\discretionary{.}{%
}{.}\hspace{.4pt}2007\hspace{.1pt}\discretionary{.}{%
}{.}\hspace{.4pt}4389006}}}


\bibitem{steinbach2000doccluster}
M.~Steinbach, G.~Karypis, and V.~Kumar.
\newblock A comparison of document clustering techniques.
\newblock 2000.

\bibitem{tu2023sdrquerier}
Y.~Tu, O.~Li, J.~Wang, H.-W. Shen, P.~Powa{\l}ko, I.~Tomescu-Dubrow, K.~M. Slomczynski, S.~Blanas, and J.~C. Jenkins.
\newblock Sdrquerier: A visual querying framework for cross-national survey data recycling.
\newblock {\em IEEE Transactions on Visualization and Computer Graphics}, 2023.

\bibitem{tu2022phrasemap}
Y.~Tu, R.~Qiu, Y.-S. Wang, P.-Y. Yen, and H.-W. Shen.
\newblock Phrasemap: Attention-based keyphrases recommendation for information seeking.
\newblock {\em IEEE Transactions on Visualization and Computer Graphics}, 2022.

\bibitem{vayansky2020review}
I.~Vayansky and S.~A. Kumar.
\newblock A review of topic modeling methods.
\newblock {\em Information Systems}, 94:101582, 2020.

\bibitem{vijaymeena2016survey}
M.~Vijaymeena and K.~Kavitha.
\newblock A survey on similarity measures in text mining.
\newblock {\em Machine Learning and Applications: An International Journal}, 3(2):19--28, 2016.

\bibitem{wang2023dataformulator}
C.~Wang, J.~Thompson, and B.~Lee.
\newblock Data formulator: Ai-powered concept-driven visualization authoring.
\newblock {\em IEEE Transactions on Visualization and Computer Graphics}, 2023.

\bibitem{wu2023llms}
T.~Wu, H.~Zhu, M.~Albayrak, A.~Axon, A.~Bertsch, W.~Deng, Z.~Ding, B.~Guo, S.~Gururaja, T.-S. Kuo, et~al.
\newblock Llms as workers in human-computational algorithms? replicating crowdsourcing pipelines with llms.
\newblock {\em arXiv preprint arXiv:2307.10168}, 2023.

\bibitem{xiang2019surveyee}
W.~Xiang and B.~Wang.
\newblock A survey of event extraction from text.
\newblock {\em IEEE Access}, 7:173111--173137, 2019.

\bibitem{yang2017vistopic}
Y.~Yang, Q.~Yao, and H.~Qu.
\newblock Vistopic: A visual analytics system for making sense of large document collections using hierarchical topic modeling.
\newblock {\em Visual Informatics}, 1(1):40--47, 2017.

\bibitem{zhang2023extractive}
H.~Zhang, X.~Liu, and J.~Zhang.
\newblock Extractive summarization via chatgpt for faithful summary generation.
\newblock {\em arXiv preprint arXiv:2304.04193}, 2023.

\bibitem{zhou2023contextfaithful}
W.~Zhou, S.~Zhang, H.~Poon, and M.~Chen.
\newblock Context-faithful prompting for large language models, 2023.

\bibitem{gilbert}
J.~Červený.
\newblock https://github.com/jakubcerveny/gilbert/commits/master generalized hilbert ("gilbert") space-filling curve for rectangular domains of arbitrary (non-power of two) sizes., 2019.

\end{thebibliography}
